\documentclass[aip,pop,reprint,amsmath,amssymb,floatfix]{revtex4-2}

\usepackage{bm}
\usepackage{booktabs}
\usepackage{graphicx}
\usepackage{xcolor}
\usepackage[
  colorlinks=true,
  linkcolor=blue,
  citecolor=blue,
  urlcolor=blue
]{hyperref}

\newcommand{\ii}{\mathrm{i}}
\newcommand{\dd}{\mathrm{d}}
\newcommand{\epsB}{\epsilon_B}
\newcommand{\kb}{K_B}
\newcommand{\ky}{k_y}

\newcommand{\omegace}{\omega_{ce}}
\newcommand{\DF}{D_F}
\newcommand{\phivec}{\bm{\phi}}
\newcommand{\orderN}{N_F}
\DeclareRobustCommand{\rev}[1]{#1}

\begin{document}

\title{Semi-local Floquet theory for active azimuthal magnetic modulation of Hall-thruster high-frequency instabilities}

\author{Yinjian Zhao}
\email{zhaoyinjian@hit.edu.cn}
\affiliation{School of Energy Science and Engineering, Harbin Institute of Technology, Harbin 150001, People's Republic of China}
\author{Changzheng Hu}
\affiliation{School of Energy Science and Engineering, Harbin Institute of Technology, Harbin 150001, People's Republic of China}

\date{\today}

\begin{abstract}
We develop a semi-local Floquet model to screen active azimuthal magnetic modulation as a control approach for high-frequency drift instabilities in Hall thrusters. The uniform-field kinetic response supplies a local kernel for a finite sideband matrix, and stability is measured by the upper growth envelope over the Bloch zone. Root searches include both frequency branches, continuation, singular-value validation, and contour root counts. At the reference normalized parallel wave number \(k_z=0.045\), for sinusoidal profiles with periods of 10, 20, and 40 mm, every nonzero sampled modulation amplitude increases both the peak and integrated positive Bloch-zone growth. The uniform-field reference is already unstable, so these increases describe faster linear amplification rather than the onset of instability from a stable state. At a period of 20 mm and relative amplitude 0.5, the peak and integral increase by 26.74 and 26.44 percent, respectively; this large amplitude requires caution within the semi-local ordering. Selected projected wave-number windows still show reductions, but these coexist with enhanced global growth and do not establish stabilization. A supplementary parallel-wave-number scan finds weak global growth reduction at \(k_z=0.09\): at 20 mm and amplitude 0.3, the peak and integral decrease by 0.300 and 0.299 percent, respectively, while the spectrum remains unstable. The sign and magnitude of the modulation effect therefore depend on the tested plasma parameters; neither these small reductions nor the local projected changes establish improved nonlinear transport.
\end{abstract}

\maketitle

\section{Introduction}
\label{sec:introduction}

Hall thrusters use a magnetic field to impede axial electron motion while ions are accelerated electrostatically. The observed cross-field electron transport often exceeds the classical collisional value, affecting discharge current, ionization, wall losses, and efficiency. Partially magnetized \(E\times B\) discharges support electrostatic fluctuations over several spatial and temporal scales; their interaction with boundaries and their contribution to anomalous transport remain active research topics. The perspective by Kaganovich \textit{et al.}\ surveys these mechanisms, experimental diagnostics, and computational challenges.\cite{Kaganovich2020EXB} The present work concerns the high-frequency, electron-drift-driven part of this spectrum.

In uniform-field kinetic theory, the relative electron--ion drift drives the electron cyclotron drift instability (ECDI), with resonant bands associated with Doppler-shifted electron cyclotron harmonics and finite-Larmor-radius effects.\cite{Ducrocq2006,Cavalier2013} A finite wave-number component along the magnetic field introduces parallel electron dynamics and permits an oblique modified two-stream instability (MTSI). Two-dimensional kinetic simulations by Janhunen \textit{et al.}\ resolve both cyclotron-harmonic growth and a longer-wavelength MTSI mode with finite parallel structure and strong parallel electron heating.\cite{Janhunen2018MTSI} We use electron drift instability (EDI) as an umbrella term for this drift-driven kinetic response. The ECDI-like and MTSI-like labels distinguish the branches of interest; small azimuthal wave number alone is not a sufficient criterion for identifying MTSI.

Kinetic simulations have established that the nonlinear spectrum and transport cannot generally be inferred from independent linear resonances. Lafleur \textit{et al.}\ connected PIC observations to a kinetic model of fluctuation-enhanced electron--ion friction,\cite{Lafleur2016Part1,Lafleur2016Part2} while Boeuf and Garrigues compared axial--azimuthal PIC results with drift-instability theory and examined cross-field transport and numerical convergence.\cite{Boeuf2018PICTheory} Janhunen \textit{et al.}\ found nonlinear transfer toward longer wavelengths in a model retaining parallel dynamics and dielectric walls.\cite{Janhunen2018MTSI} Comparisons of PIC and continuum Vlasov simulations by Tavassoli \textit{et al.}\ showed that azimuthal domain length and particle noise affect resolved growth rates, nonlinear spectral evolution, heating, and current.\cite{Tavassoli2023PICVlasov} More recently, Bello-Ben\'{i}tez \textit{et al.}\ demonstrated that particle injection and residence time strongly affect whether nonlinear ECDI oscillations and cross-field current persist.\cite{BelloBenitez2025Injection} Three-dimensional simulations also reveal coupling between ECDI and ion--ion two-stream instability.\cite{Denig2023POP,Zhao2026PSTReview3DPIC} These findings distinguish linear amplification from the nonlinear saturation and fluctuation correlations that determine transport.

Experiments provide complementary constraints on the kinetic models. Probe measurements reconstructed high-frequency Hall-thruster dispersion relations.\cite{Lazurenko2008POP} Using nonintrusive collective light scattering, Tsikata \textit{et al.}\ measured anisotropic small-scale density fluctuations and dispersion relations along the thruster axis and the \(E\times B\) direction.\cite{Tsikata2009Scattering} The measured \(E\times B\) branch had approximately ion-acoustic-like dispersion, without resolved discrete cyclotron-harmonic bands over the sampled wave-number interval. Thus, agreement in characteristic frequency and wave-number scales did not establish agreement with every feature of the local linear spectrum. Cavalier \textit{et al.}\ subsequently compared kinetic dispersion calculations with measured fluctuation spectra,\cite{Cavalier2013} and Brown and Jorns directly measured instability growth and nonlinear energy transfer and used measured growth rates to improve wave-driven transport estimates.\cite{Brown2023PRL,Brown2023PRE} Together, these studies motivate a linear spectral model whose assumptions and relation to nonlinear transport are stated explicitly.

The high-frequency response studied here is only part of the instability spectrum. Long-wavelength gradient-drift modes, including the collisionless or modified Simon--Hoh instability, can have low real frequencies without correspondingly weak growth. Hara, Mansour, and Tsikata analyze density-gradient-driven modes in a partially magnetized plasma, including a modified Simon--Hoh branch.\cite{Hara2022GradientDrift} Romadanov \textit{et al.}\ show that nonlocal equilibrium profiles produce multiple gradient-drift eigenmodes and stability predictions that differ from local criteria, particularly at small azimuthal wave number.\cite{Romadanov2016Nonlocal} Such modes can be important for transport when their spatial extent samples background inhomogeneity. Their growth need not be negligible compared with high-frequency branches; a distinction in real frequency alone does not establish a hierarchy of growth rates. The present calculation does not include the gradient-driven response needed to assess these modes.

We use dimensionless frequencies, wave numbers, coordinates, and velocities as defined in Sec.~\ref{sec:uniform_response}; dimensional counterparts are explicitly identified by the superscript ``phys''. The magnetic field is therefore not merely a background parameter in the instability problem. In a crossed-field plasma it sets the electron cyclotron frequency \(\omegace\), the electron Larmor radius \(\rho_e\), the \(E\times B\) drift speed \(V_E\), and the approximate electron-cyclotron resonance locations \(k_y V_E\simeq n\omegace\). A change in magnetic-field strength, gradient, or spatial periodicity can consequently move resonant wave numbers, modify finite-Larmor-radius weights, and alter the relative importance of low-\(k_y\) MTSI-like response and higher-\(k_y\) ECDI-like response. Recent radial--azimuthal PIC calculations of radial magnetic-field strength and gradient effects indicate that tailored magnetic profiles can suppress EDI development and anomalous axial transport by changing the relative electron-ion drift, electron confinement, and unstable spectral range.\cite{Chen2026PSTRadialB} This sensitivity makes magnetic-field shaping a natural candidate for instability control, but it also makes the response difficult to infer from geometry alone.

Azimuthal magnetic-field nonuniformity can arise from the magnetic-circuit design and, in practice, from manufacturing or assembly imperfections. In SPT-type Hall thrusters, magnetic poles and conductive columns can introduce azimuthal variations of the radial magnetic field. These departures from an ideal azimuthally uniform field motivated our investigation of how magnetic nonuniformity affects high-frequency instabilities. The related two-dimensional PIC studies by Zhou \textit{et al.}\ examined azimuthally inhomogeneous magnetic fields and their effects on the instability spectrum and cross-field transport in radial--azimuthal and axial--azimuthal geometries.\cite{Zhou2025PSST,Zhou2026PSSTAxial}

Recent two-dimensional radial--azimuthal PIC simulations showed that this passive azimuthal magnetic-field inhomogeneity can modify the instability through competing mechanisms.\cite{Zhou2025PSST} The magnetic variation modulates the azimuthal electron drift velocity, which can strengthen the local drift-instability drive, while it also broadens the EDI wave-number spectrum and reduces discrete resonant peaks. In that radial--azimuthal model, inhomogeneity levels below \(5\%\) produced relatively small changes, whereas a \(10\%\) inhomogeneity reduced the ECDI saturation intensity by \(13.4\%\) and the electron mobility by \(17.4\%\). Complementary axial--azimuthal PIC results were reported by Zhou \textit{et al.}\cite{Zhou2026PSSTAxial} This study evaluates the effective axial electron mobility from the fluctuation-induced transport term. For inhomogeneity levels up to \(5\%\), it reports a maximum enhancement of \(12.1\%\) relative to the uniform-field case, with a local dip at \(5\%\). At \(10\%\), the enhancement reaches \(37.3\%\). The enhanced transport is associated with broadening around the ECDI branch and stronger neighboring modes; the instability intensity also responds to the self-consistently formed axial electric field. These values refer specifically to the effective-mobility diagnostic in the journal article. Taken together, these simulations show that magnetic inhomogeneity is physically important, but its effect is geometry-dependent and mixed with nonlinear saturation, wall coupling, and transport closure.

These observations motivate asking whether an azimuthal magnetic variation can be deliberately selected to modify, and preferably reduce, the unstable spectrum. The present work treats its amplitude and wavelength as independently prescribed parameters so that the linear effect can be assessed before a candidate configuration is investigated in experiments or extensive PIC calculations. Given a prescribed sinusoidal azimuthal modulation,
\begin{equation}
  B(y)=B_0[1+\epsB\cos(\kb y)],
  \label{eq:intro_by_modulated}
\end{equation}
\rev{with independently chosen amplitude \(\epsB\) and modulation wave number \(\kb=2\pi\lambda_D/L_B\) with \(y=y^{\rm phys}/\lambda_D\) and dimensional period \(L_B\), which parts of the high-frequency drift-instability spectrum can be linearly reduced, and under what validity limits?}

Direct PIC simulation is ultimately needed to validate any promising magnetic-control scenario, because it contains nonlinear saturation, density redistribution, wall effects, and fluctuation-driven transport. It is not, however, an economical first tool for scanning a two-parameter space of modulation amplitude and wavelength. Fully three-dimensional PIC studies are especially valuable for resolving axial acceleration, radial wall/sheath coupling, and azimuthal drift dynamics, but their cost motivates careful linear pre-screening before large parameter scans are attempted.\cite{Zhao2026PSTReview3DPIC} PIC results also mix the linear spectral response with nonlinear mode coupling and boundary-mediated transport, making it difficult to identify whether a given modulation has a direct linear stabilizing tendency.

A uniform-field kinetic EDI dispersion relation provides a useful starting point for such a screening model.\cite{Ducrocq2006,Lafleur2016Part1,Lafleur2016Part2} In a uniform magnetic field, the unperturbed electron orbit has the standard \(E\times B\) drift plus cyclotron motion, and the linear susceptibility retains the finite-Larmor-radius, cyclotron-harmonic, Doppler-shift, and Gordeev-function structure needed to describe the broad EDI spectrum, including its cyclotron-resonant and low-\(k_y\) branches. The numerical implementation and fitting strategy used by Cavalier \textit{et al.}\ further demonstrate the practical value of this scalar dispersion framework for Hall-thruster fluctuation analysis.\cite{Cavalier2013} Its limitation for the present problem is equally clear: in a uniform field, each Fourier mode \(k_y\) is independent, whereas a periodic magnetic modulation couples \(k_y\) to \(k_y\pm \kb\), \(k_y\pm2\kb\), and higher sidebands.

For the prescribed periodic linear operator, a Bloch--Floquet expansion organizes this coupling into sidebands separated by \(\kb\). The specific contribution sought here is a tractable extension of the uniform-field kinetic response to these linearly coupled sidebands. This coupling is induced by the imposed magnetic periodicity and is distinct from the nonlinear interactions studied in the kinetic simulations reviewed above. The resulting eigenfrequencies describe linear spectral changes; they do not determine saturated fluctuation amplitudes or electron mobility.

A complete kinetic theory in the nonuniform field \(B(y)\) would require solving the unperturbed electron orbits in the spatially varying magnetic field and reintegrating the linearized Vlasov response along those orbits. Along such trajectories, \(\omegace(y)\), \(\rho_e(y)\), and \(V_E(y)\) all vary, so the simple uniform-orbit Bessel and cyclotron-harmonic expansion is no longer directly available. A full nonuniform-orbit derivation lies beyond the present approximation; the screening results below remain conditional on using the uniform response as a local kernel.

The present work therefore constructs a semi-local Floquet coupled-mode model. The uniform-field kinetic EDI response is retained as a local spectral kernel, while the imposed periodic magnetic modulation is represented by coupling among Floquet sidebands \(k_{ym}=q+m\kb\), where \(q\) is the Bloch quasi-wavenumber and \(m\) is an integer sideband index. Here \(D_0\) denotes the scalar uniform-field dispersion residual, with the fixed \(k_x\) and \(k_z\) arguments suppressed, and \(\DF\) denotes the finite Floquet dispersion matrix coupling the retained Fourier sidebands. Their explicit forms are given in Secs.~\ref{sec:uniform_response} and \ref{sec:matrix_relation}, respectively. The scalar condition
\begin{equation}
  D_0(\omega,k_y;B_0)=0
\end{equation}
is thereby replaced by the matrix dispersion relation
\begin{equation}
  \det \DF(\omega,q;\epsB,\kb)=0.
\end{equation}
This construction is not a transport closure and not a substitute for full nonuniform kinetic theory. It is a linear pre-screening model for deciding whether a prescribed magnetic modulation has a direct tendency to reduce the high-frequency growth spectrum.

An important practical issue is how suppression should be measured in a folded Floquet spectrum. A single Floquet root contains several sidebands, so assigning that root to one projected \(k_y\) can produce apparent gaps, jumps, or local reductions when the dominant sideband changes. The primary stability diagnostic used here is therefore the Bloch-zone growth envelope
\begin{equation}
  \Gamma(q)=\max_j \operatorname{Im}\omega_j(q),
\end{equation}
where \(j\) indexes the accepted roots at fixed \(q\). Here and below, ``global'' refers to the represented model spectrum across the Bloch zone, rather than to all instability mechanisms of an inhomogeneous discharge. We also examine how a mode is distributed among the coupled spatial harmonics and check how accurately each computed frequency satisfies the matrix equation. These diagnostics are defined in Secs.~\ref{sec:floquet_root_method} and \ref{sec:numerical_results}; neither a change in mode structure nor a small numerical residual alone establishes stabilization.

The calculation quantifies changes in linear growth for a prescribed sinusoidal magnetic profile. Here a growth metric means a scalar quantitative measure, specifically the peak of the growth-rate envelope or its positive integral across the Bloch zone. In the audited period scans at the reference \(k_z=0.045\) and \(L_B=10,20,40~\mathrm{mm}\), every sampled nonzero amplitude raises both the peak and integrated positive Bloch-zone envelope relative to the already unstable uniform-field reference. Selected projected sideband windows can nevertheless show reduced growth measures. We distinguish these local changes from global suppression, and restrict the physical interpretation using the semi-local ordering. The additional parallel-wave-number scan in Sec.~\ref{sec:kz_scan} finds weak global growth reduction at \(k_z=0.09\), while the other three sampled values show enhancement. Thus the reference-case trend does not apply to every choice of plasma parameters. Section~\ref{sec:theory} formulates the semi-local Floquet model, Sec.~\ref{sec:numerical_method} describes the root search and checks, Sec.~\ref{sec:numerical_results} presents the scans, and Sec.~\ref{sec:conclusions} discusses the implications for magnetic-modulation screening.

\section{Semi-local Floquet theory for periodically modulated magnetic fields}
\label{sec:theory}

\subsection{Purpose and level of approximation}
\label{sec:theory_scope}

The theoretical objective of this work is to construct a tractable linear model for how a prescribed azimuthal magnetic-field modulation modifies the high-frequency drift-instability spectrum of a Hall-thruster plasma. The model is not presented as a complete kinetic theory for a fully nonuniform crossed-field equilibrium. Rather, it is a semi-local coupled-mode extension of the uniform-field kinetic dispersion relation. The uniform problem supplies the local wave-particle response of magnetized electrons, while the periodic magnetic modulation supplies the Fourier coupling that mixes neighboring azimuthal sidebands.

The starting point is a uniform-field kinetic dispersion condition \(D_0(\omega,\ky;B_0)=0\), where \(D_0\) is the scalar dispersion residual, \(\omega\) is the complex angular frequency, \(k_y\) is the azimuthal wave number, and \(B_0\) is the uniform reference magnetic-field strength. In this uniform problem, the unperturbed electron orbit is known analytically, and the electron susceptibility contains the usual finite-Larmor-radius and cyclotron-harmonic structure.\cite{Ducrocq2006,Cavalier2013} In the present problem the imposed field is periodic in the azimuthal coordinate \(y\), as specified in Eq.~\eqref{eq:intro_by_modulated}. Here \(B(y)\) is the prescribed magnetic-field strength, \(\epsB\) is its relative modulation amplitude, and \(K_B\) is the modulation wave number.
The model addresses a limited question: if the uniform kinetic response \(D_0(\omega,\ky;B_0)=0\) is periodically modulated through Eq.~\eqref{eq:intro_by_modulated}, how are the Fourier components \(\ky\), \(\ky\pm\kb\), \(\ky\pm2\kb\), and so on coupled, and how does that sideband coupling modify the linear eigenfrequency?

The resulting theory should therefore be interpreted as a semi-local Floquet coupled-mode model based on the uniform-field kinetic EDI response. The qualifier ``semi-local'' specifies the scope of the approximation. The calculation keeps the uniform-field kinetic response as a spectral building block, but it does not recompute the exact electron orbits in \(B(y)\), does not solve a self-consistent nonuniform equilibrium distribution \(f_0(y,\bm{v})\), and does not determine nonlinear saturation or anomalous electron mobility. Here \(f_0\) denotes the equilibrium distribution function and \(\bm{v}\) is particle velocity. These limitations define the intended use of the theory as a linear pre-screening model.

\subsection{Uniform-field kinetic response}
\label{sec:uniform_response}

The normalization is based on a singly charged, quasineutral reference plasma with density \(n_0\), electron temperature \(T_e\) expressed in energy units, ion mass \(m_i\), and electron mass \(m_e\). The reference scales are
\begin{equation}
  \begin{gathered}
  \omega_{pi}=\sqrt{\frac{n_0e^2}{\varepsilon_0m_i}},\qquad
  \lambda_D=\sqrt{\frac{\varepsilon_0T_e}{n_0e^2}},\\
  c_s=\lambda_D\omega_{pi}=\sqrt{\frac{T_e}{m_i}},
  \end{gathered}
  \label{eq:normalization_scales}
\end{equation}
where \(\varepsilon_0\) is the vacuum permittivity. With the superscript ``phys'' denoting a dimensional quantity, the dimensionless variables are
\begin{equation}
  \begin{aligned}
    \widehat\omega&=\omega^{\rm phys}/\omega_{pi},&
    \widehat\gamma&=\gamma^{\rm phys}/\omega_{pi},\\
    \widehat k_\alpha&=k_\alpha^{\rm phys}\lambda_D,&
    \widehat q&=q^{\rm phys}\lambda_D,\\
    \widehat K_B&=K_B^{\rm phys}\lambda_D,&
    \widehat y&=y^{\rm phys}/\lambda_D,\\
    \widehat V&=V^{\rm phys}/c_s,&
    \widehat\omega_{ce}&=\omega_{ce}^{\rm phys}/\omega_{pi},\\
    \widehat\rho_e&=\rho_e^{\rm phys}/\lambda_D,&
    M&=m_i/m_e .
  \end{aligned}
  \label{eq:normalized_variables}
\end{equation}
Here \(\alpha=x,y,z\), \(\omega_{ce}^{\rm phys}=eB/m_e\), and the reference value is \(\omega_{ce0}=(eB_0/m_e)/\omega_{pi}\). The velocity mapping applies to \(V_d,V_p,V_i,V_E\), particle and drift velocities, and the reference thermal speed \(v_{the}\). The other coordinates are scaled as \(y\), and \(\widehat t=\omega_{pi}t^{\rm phys}\). Hats are omitted after displaying the mapping: throughout the model equations, numerical parameters, and appendices, \(\omega,\gamma,\bm{k},q,K_B,y,t\), velocities, and \(\rho_e\) denote dimensionless variables. Their dimensional counterparts carry the superscript ``phys'', including in the orbit discussion of Sec.~\ref{sec:full_nonuniform_difficulty}. Thus the same normalized wave numbers occur in the scalar residual, kinetic arguments, differential operator, and Floquet sidebands.

The reference scales \(\omega_{pi},\lambda_D,c_s\), the dimensional background quantities \(n_0,T_e,B_0,\bm{E}_0\), and the physical modulation period \(L_B\) retain their units. The magnetic-field strength \(B\) enters the normalized response through \(s=B/B_0\); \(\epsB\) is dimensionless. The period in the normalized coordinate is \(\ell_B=L_B/\lambda_D\), so \(K_B=2\pi/\ell_B=2\pi\lambda_D/L_B\) and \(K_B^{\rm phys}=2\pi/L_B\). Consequently, \(\chi=\epsB K_B\rho_e=\epsB K_B^{\rm phys}\rho_e^{\rm phys}\). All reference scales remain fixed as the field is modulated; in particular, \(c_s=\lambda_D\omega_{pi}\). Growth envelopes \(\Gamma\), their absolute differences, and \(I_q\) are dimensionless; their definitions below use the normalized frequency. Figure captions explicitly identify any retained axis notation that displays a dimensional quantity divided by its reference scale.

The one-component Maxwellian thermal speed is \(v_{the}^{\rm phys}=\sqrt{T_e/m_e}\), giving \(v_{the}=\sqrt{M}\) before rounding. The factor \(\sqrt{2}\) enters the plasma-dispersion-function argument in Eq.~\eqref{eq:gordeev_series}. The reference electron thermal Larmor radius is \(\rho_e^{\rm phys}=v_{the}^{\rm phys}/\omega_{ce0}^{\rm phys}\), or \(\rho_e=\rho_e^{\rm phys}/\lambda_D=v_{the}/\omega_{ce0}\). It characterizes the finite-Larmor-radius response and the ordering estimate below. Wave numbers remain normalized by \(\lambda_D\), not by \(\rho_e^{\rm phys}\). The independently rounded thermal-speed and mass-ratio inputs used in the calculations are specified in Sec.~\ref{sec:ion_rotation_method}. For reference, we write the local uniform-field response in the normalized scalar form used by the solver. With fixed \(k_x\), \(k_y\), and \(k_z\), the scalar residual can be written as
\begin{equation}
  D_0(\omega,k_x,k_y,k_z;B_0)
  =
  (\omega-k_xV_p)^2
  -
  \frac{k^2}{1+k^2+g(\Omega,X,Y)},
  \label{eq:scalar_kinetic_residual}
\end{equation}
where
\(k^2=k_x^2+k_y^2+k_z^2\), and
\begin{equation}
  \Omega=\frac{\omega-k_yV_d}{\omegace},\qquad
  X=\frac{(k_x^2+k_y^2)M}{\omegace^2},\qquad
  Y=\frac{k_z^2M}{\omegace^2}.
  \label{eq:gordeev_args}
\end{equation}
Here every \(k_\alpha\) in Eqs.~\eqref{eq:scalar_kinetic_residual} and \eqref{eq:gordeev_args} means \(k_\alpha^{\rm phys}\lambda_D\); in particular, \(k^2=\lambda_D^2|\bm{k}^{\rm phys}|^2\). The frequencies and velocities in both equations are also normalized according to Eq.~\eqref{eq:normalized_variables}. Here \(V_d\) is the electron \(E\times B\) drift velocity, \(V_p\) is the ion axial velocity in the dimensionless normalization used here, \(M=m_i/m_e\) is the ion-to-electron mass ratio, and \(g\) is the Gordeev function. In the cyclotron-harmonic representation used below,
\begin{equation}
  g(\Omega,X,Y)
  =
  \frac{\Omega}{\sqrt{2Y}}
  \sum_{p=-\infty}^{\infty}
  \Gamma_p(X)
  Z\!\left(
  \frac{\Omega-p}{\sqrt{2Y}}
  \right),
  \label{eq:gordeev_series}
\end{equation}
where \(p\) is the integer electron-cyclotron harmonic index, \(Z(z)=\ii\sqrt{\pi}\,e^{-z^2}\operatorname{erfc}(-\ii z)\) is the Landau-continued plasma dispersion function, and \(\Gamma_p(X)=e^{-X}I_p(X)\) is the finite-Larmor-radius Bessel weight of harmonic \(p\). This series representation is used for \(Y>0\), as in every reported case; \(k_z=0\) requires taking the corresponding limit. Here \(I_p\) is the modified Bessel function of the first kind. The quantity \(\omegace\) is the electron cyclotron angular frequency divided by \(\omega_{pi}\). Since \(c_s=\lambda_D\omega_{pi}\), the ratio \(\Omega=(\omega-k_yV_d)/\omegace\) equals \((\omega^{\rm phys}-k_y^{\rm phys}V_d^{\rm phys})/\omega_{ce}^{\rm phys}\); no dimensional wave number is inserted into the normalized residual. Since \(T_e/m_e=M c_s^2\), the other two arguments equivalently read
\begin{equation*}
  \begin{aligned}
    X&=\frac{[(k_x^{\rm phys})^2+(k_y^{\rm phys})^2](T_e/m_e)}{(\omega_{ce}^{\rm phys})^2},\\
    Y&=\frac{(k_z^{\rm phys})^2(T_e/m_e)}{(\omega_{ce}^{\rm phys})^2}.
  \end{aligned}
\end{equation*}
These identities account for the factor \(M\) in Eq.~\eqref{eq:gordeev_args}. The arguments \(\Omega\), \(X\), and \(Y\) are dimensionless inputs to the Gordeev function: \(\Omega\) is the electron-frame frequency normalized by \(\omegace\), \(X\) measures perpendicular finite-Larmor-radius response, and \(Y\) measures the parallel thermal response. The complex root \(\omega=\omega_r+\ii\gamma\) gives the real angular frequency \(\omega_r\) and linear growth rate \(\gamma\); \(\ii=\sqrt{-1}\). A positive \(\gamma\) indicates temporal growth.

Equation~\eqref{eq:scalar_kinetic_residual} has a specific role in the present paper. It is not rederived here from the Vlasov-Poisson system. Instead, it is treated as the verified uniform-field kinetic kernel from which the semi-local Floquet problem is built. The finite-Larmor-radius dependence enters through \(X\), the parallel thermal phase mixing enters through \(Y\), and the cyclotron-resonant structure enters through the argument \(\Omega\). Thus, even though the Floquet extension is approximate, it retains the central local kinetic physics of the high-frequency electron drift instability.

\subsection{Difficulty of the full nonuniform kinetic problem}
\label{sec:full_nonuniform_difficulty}

The uniform-field derivation is analytically possible because the unperturbed electron motion in constant crossed fields has a simple structure: a constant \(E\times B\) drift superposed on circular gyromotion. This permits the linearized Vlasov equation to be integrated along unperturbed characteristics and reduced to a Bessel-function and cyclotron-harmonic response.

That simplification is lost when the imposed magnetic field varies in the azimuthal direction. For this orbit discussion, dimensional variables are explicitly marked by ``phys''. Write \(B_{\rm phys}(y^{\rm phys})\equiv B(y^{\rm phys}/\lambda_D)\), where the subscript specifies the spatial argument; the field strength itself retains its units in either notation. For \(\bm{B}=B_{\rm phys}(y^{\rm phys})\hat{\bm{z}}\), the unperturbed orbit satisfies schematically
\begin{equation}
  \begin{aligned}
  \frac{\dd\bm{r}^{\rm phys}}{\dd t^{\rm phys}}&=\bm{v}^{\rm phys},\\
  m_e\frac{\dd\bm{v}^{\rm phys}}{\dd t^{\rm phys}}
  &=-e[\bm{E}_0+\bm{v}^{\rm phys}\times B_{\rm phys}(y^{\rm phys})\hat{\bm{z}}].
  \end{aligned}
  \label{eq:nonuniform_orbit}
\end{equation}
Along such an orbit,
\begin{equation*}
  \begin{aligned}
  \omega_{ce}^{\rm phys}(y^{\rm phys})&=\frac{eB_{\rm phys}(y^{\rm phys})}{m_e},\\
  \rho_e^{\rm phys}(y^{\rm phys})&=\frac{v_\perp^{\rm phys}}{\omega_{ce}^{\rm phys}(y^{\rm phys})},\\
  V_E^{\rm phys}(y^{\rm phys})&=\frac{E_0}{B_{\rm phys}(y^{\rm phys})}.
  \end{aligned}
\end{equation*}
are no longer constants. Here \(\bm{B}\) is the magnetic-field vector, \(\hat{\bm{z}}\) is the unit vector along the field direction, \(\bm{r}^{\rm phys}\) is dimensional particle position, \(m_e\) is the electron mass, \(e\) is the elementary charge, and \(\bm{E}_0\) is the imposed background electric field with magnitude \(E_0\). The quantities \(\omega_{ce}^{\rm phys}\), \(\rho_e^{\rm phys}\), and \(V_E^{\rm phys}\) are the dimensional local electron gyrofrequency, Larmor radius, and \(E\times B\) drift speed, respectively; \(v_\perp^{\rm phys}\) denotes the perpendicular gyration speed relative to the local guiding-center drift. In the nonuniform field, \(\rho_e^{\rm phys}(y^{\rm phys})\) is a local gyro-radius estimate; it is not an exact circular-orbit radius. The phase factor that appears in the orbit integral cannot be reduced to the same closed-form Bessel expansion as in the uniform theory. A rigorous nonuniform-orbit kinetic theory would require solving or approximating Eq.~\eqref{eq:nonuniform_orbit} and then reintegrating the linearized Vlasov equation along those nonuniform characteristics. That calculation is substantially more difficult and is not attempted in the present work.

Magnetic inhomogeneity can alter the response beyond a local change of cyclotron frequency. Ramos and Ripoli derive a drift-kinetic electron response with cold fluid ions, retaining the kinetic resonance associated with magnetic-gradient drift and obtaining modified instability thresholds.\cite{RamosRipoli2026Kinetic} Their theory concerns low-frequency, long-perpendicular-wavelength modes with no parallel dynamics; its dispersion relation cannot be directly substituted for the cyclotron-harmonic kernel used here. In a fluid treatment, Frias \textit{et al.}\ show that accounting for electron-flow compressibility in an inhomogeneous magnetic field changes instability criteria.\cite{Frias2012GradientFluid} These studies establish that magnetic gradients can introduce additional dynamics, whose relevance depends on the geometry and ordering.

The present calculation describes sideband coupling of a uniform-field kinetic response. It does not include the complete magnetic-gradient drift response, deformation of the unperturbed orbits, or self-consistent changes of the equilibrium plasma profiles. The geometric restriction on one direct magnetic-drift coupling in our scans is discussed in Sec.~\ref{sec:validity}; it does not remove all of these limitations.

\subsection{Semi-local operator model}
\label{sec:semilocal_operator_model}

Although the full nonuniform kinetic problem is difficult, the uniform dispersion relation remains useful because it contains the essential local wave-particle response of magnetized electrons: finite-Larmor-radius effects, Doppler shifting by the electron drift, and resonances with electron cyclotron harmonics. For weak or moderate prescribed modulation, the model treats \(D_0(\omega,\ky;B)\) as a local spectral response function evaluated at the local magnetic field.

The compact semi-local modeling step is
\begin{equation}
  D_0(\omega,\ky;B_0)=0
  \quad\longrightarrow\quad
  D_0(\omega,-\ii\partial_y;B(y))\phi(y)=0.
  \label{eq:operator_replacement}
\end{equation}
In a uniform system \(k_y\) is a scalar label because each Fourier component evolves independently. In a periodic system \(k_y\) must be represented by the differential operator \(-\ii\partial_y\), acting on the full azimuthal eigenfunction \(\phi(y)\). Both \(k_y\) and \(y\) in Eq.~\eqref{eq:operator_replacement} are dimensionless: \(-\ii\partial_y=\lambda_D(-\ii\partial_{y^{\rm phys}})\), consistent with \(k_y=\lambda_D k_y^{\rm phys}\). The replacement is the usual Fourier correspondence between wave number and spatial derivative, but its consequence is important: once \(B(y)\) is periodic, the operator in Eq.~\eqref{eq:operator_replacement} has periodic coefficients and couples sidebands separated by \(\kb\).

Equation~\eqref{eq:operator_replacement} is a semi-local modeling approximation. It retains the local dependence of the uniform susceptibility on magnetic-field strength, including changes in the cyclotron frequency, Larmor radius, \(E\times B\) drift, and resonance locations. Periodicity then couples the Fourier components. These retained effects are distinct from the explicit gradient-dependent terms and orbit corrections obtained by deriving the response in a nonuniform equilibrium. No quantitative bound on those omitted corrections is established here. The detailed meaning of the operator replacement is given in Appendix~\ref{app:ky_operator}.

\subsection{Magnetic-field expansion and first-order coupling}
\label{sec:field_expansion}

For small modulation amplitude, the semi-local operator can be expanded about the uniform reference field. Write
\begin{equation*}
  B(y)=B_0+\delta B(y),\qquad
  \delta B(y)=\epsB B_0\cos(\kb y).
\end{equation*}
Then
\begin{widetext}
\begin{equation}
  D_0(\omega,-\ii\partial_y;B(y))\phi(y)
  \simeq
  \left[
    D_0(\omega,-\ii\partial_y;B_0)
    +
    \delta B(y)
    \left.
    \frac{\partial D_0(\omega,-\ii\partial_y;B)}{\partial B}
    \right|_{B_0}
  \right]\phi(y).
  \label{eq:operator_taylor}
\end{equation}
\end{widetext}
The first term is the uniform-field response. The second term is the lowest-order correction produced by the periodic magnetic modulation. Higher-order terms in \(\delta B\) would produce couplings to more distant sidebands and nonlinear-in-\(\epsB\) corrections to the matrix elements. In the present theory, only the first-order term is retained.

It is often cleaner in numerical work to scale the field by \(s=B/B_0\). Then \(s(y)=1+\epsB\cos(\kb y)\), and the response is expanded as
\begin{equation}
  D_0(\omega,\ky;s)
  \simeq
  D_0(\omega,\ky;1)
  +
  [s(y)-1]
  \left.
  \frac{\partial D_0(\omega,\ky;s)}{\partial s}
  \right|_{s=1}.
  \label{eq:s_expansion}
\end{equation}
For fixed imposed electric field, changing \(B\) changes both the cyclotron frequency and the electron drift: \(\omegace(s)=s\,\omega_{ce0}\) and \(V_d(s)=V_{d0}/s\). Here \(s\) is the dimensionless magnetic-field scale, \(\omega_{ce0}\) is the reference electron cyclotron angular frequency at \(B_0\), and \(V_{d0}\) is the reference electron \(E\times B\) drift at \(B_0\). Thus the magnetic derivative in Eq.~\eqref{eq:s_expansion} contains both resonance detuning through \(\omegace\) and drift modification through \(V_d\). The derivative holds \(\omega,\bm{k},n_0,T_e,\bm{E}_0\), the imposed ion velocities, and all reference scales fixed. Because \(B=B_0s\), \(B_0\partial_B D_0=\partial_sD_0\); hence \((\epsB B_0/2)\partial_BD_0=(\epsB/2)\partial_sD_0\). No additional frequency or length scale multiplies the normalized matrix entries.

\subsection{Floquet representation}
\label{sec:floquet_representation}

The field in Eq.~\eqref{eq:intro_by_modulated} is invariant under translations by the normalized period \(\ell_B=2\pi/\kb=L_B/\lambda_D\). Here \(L_B\) is the dimensional spatial period of the imposed magnetic modulation. The continuous translational symmetry of the uniform system is therefore replaced by discrete translational symmetry. For any linear operator with periodic coefficients, the eigenfunction can be written in Floquet form,
\begin{equation}
  \phi(y)=e^{\ii qy}u_q(y),\qquad
  u_q(y+\ell_B)=u_q(y),
  \label{eq:floquet_form}
\end{equation}
where \(q\) is the Floquet quasi-wavenumber and \(u_q(y)\) is the periodic part of the eigenfunction associated with \(q\). Since \(u_q\) is periodic, it has the Fourier expansion
\begin{equation*}
  u_q(y)=\sum_{\ell=-\infty}^{\infty}\phi_\ell e^{\ii \ell\kb y}.
\end{equation*}
The coefficient \(\phi_\ell\) is the complex amplitude of the \(\ell\)-th Floquet harmonic.
Therefore,
\begin{equation}
  \phi(y)=
  \sum_{\ell=-\infty}^{\infty}
  \phi_\ell e^{\ii(q+\ell\kb)y}.
  \label{eq:floquet_series}
\end{equation}
The normalized sideband wave numbers are \(k_{y\ell}=q+\ell\kb\), corresponding to \(k_{y\ell}^{\rm phys}=(q+\ell\kb)/\lambda_D\), with \(\ell=0,\pm1,\pm2,\ldots\), and the quasi-wavenumber may be restricted to the first Brillouin zone \(-\kb/2<q\le\kb/2\). The phase \((q+\ell K_B)y\) therefore equals \(k_{y\ell}^{\rm phys}y^{\rm phys}\). Thus a single eigenmode in the periodically modulated system is not a single Fourier component. It is a coupled packet of sidebands separated by integer multiples of \(\kb\). Appendix~\ref{app:floquet_waves} gives the translational-symmetry derivation of Eq.~\eqref{eq:floquet_form}.

\subsection{Floquet matrix dispersion relation}
\label{sec:matrix_relation}

Substituting Eq.~\eqref{eq:floquet_series} into the semi-local operator equation and projecting onto the sideband basis gives the coupled sideband system \(\sum_m D_{\ell m}(\omega,q)\phi_m=0\). All \(q,K_B,k_{ym}=q+mK_B\) in this matrix use the same normalized units as the scalar residual. Here \(D_{\ell m}\) is the matrix element coupling input sideband \(m\) to output sideband \(\ell\), and \(\phi_m\) is the amplitude of the input sideband. At first order in \(\epsB\), the matrix elements are
\begin{widetext}
\begin{equation}
  D_{\ell m}(\omega,q)
  =
  D_0(\omega,k_{ym};B_0)\delta_{\ell m}
  +
  \frac{\epsB B_0}{2}
  \left.
  \frac{\partial D_0(\omega,k_{ym};B)}{\partial B}
  \right|_{B_0}
  \left(\delta_{\ell,m+1}+\delta_{\ell,m-1}\right).
  \label{eq:matrix_element_b}
\end{equation}
\end{widetext}
Equivalently, using the dimensionless field scale \(s=B/B_0\),
\begin{widetext}
\begin{equation}
  D_{\ell m}(\omega,q)
  =
  D_0(\omega,k_{ym};1)\delta_{\ell m}
  +
  \frac{\epsB}{2}
  \left.
  \frac{\partial D_0(\omega,k_{ym};s)}{\partial s}
  \right|_{s=1}
  \left(\delta_{\ell,m+1}+\delta_{\ell,m-1}\right).
  \label{eq:matrix_element_s}
\end{equation}
\end{widetext}
The symbol \(\delta_{\ell m}\) is the Kronecker delta, equal to one when \(\ell=m\) and zero otherwise. The diagonal term is the uncoupled uniform-field response of each sideband. The off-diagonal term is the nearest-neighbor sideband coupling produced by the sinusoidal magnetic modulation. The factor \(1/2\) comes from \(\cos(\kb y)=[e^{\ii\kb y}+e^{-\ii\kb y}]/2\), and the sideband selection rule follows from
\begin{align*}
  \cos(\kb y)e^{\ii(q+m\kb)y}
  &=
  \frac{1}{2}e^{\ii[q+(m+1)\kb]y}
  \\
  &\quad+
  \frac{1}{2}e^{\ii[q+(m-1)\kb]y}.
\end{align*}
Therefore, a first-harmonic magnetic modulation couples the \(m\)-th input sideband only to the neighboring output sidebands \(m+1\) and \(m-1\). Appendix~\ref{app:projection} gives the projection integral and the Kronecker-delta selection rule explicitly.

After finite truncation to \(m=-\orderN,\ldots,\orderN\), the sideband system can be written as \(\DF(\omega,q;\epsB,\kb)\phivec=0\). Here \(\DF\) is the finite Floquet matrix built from the scalar kinetic kernel, \(\orderN\) is the sideband half-width retained in the truncated basis, and \(\phivec=(\phi_{-\orderN},\ldots,\phi_{\orderN})^{\mathsf{T}}\) is the corresponding sideband-amplitude vector. A nontrivial sideband vector requires
\begin{equation}
  \det \DF(\omega,q;\epsB,\kb)=0.
  \label{eq:det_condition}
\end{equation}
This is the Floquet matrix dispersion relation. In the limit \(\epsB=0\), all off-diagonal couplings vanish, and the matrix reduces to a folded set of independent uniform-field scalar dispersion relations. Appendix~\ref{app:matrix_dispersion} expands this scalar-to-matrix transition and gives a minimal three-sideband example.

\subsection{Validity conditions and interpretation}
\label{sec:validity}

The model is expected to be most reliable under the following conditions. For \(B_0>0\) and \(0\le\epsB<1\), the prescribed sinusoidal profile already satisfies \(B(y)\ge B_0(1-\epsB)>0\); a separate no-field-reversal condition is therefore unnecessary. The first-order expansion about \(B_0\) additionally requires weak or moderate modulation: positivity alone does not establish the accuracy of this expansion. The field should also vary only weakly over a gyro-orbit. A useful practical estimate is
\begin{equation}
  \chi=\epsB\kb\rho_e\ll1.
  \label{eq:chi_condition}
\end{equation}
Here \(\chi\) is the semi-local ordering parameter and \(\rho_e\) is the reference electron thermal Larmor radius defined in Sec.~\ref{sec:uniform_response}. This condition expresses that the magnetic field should not vary substantially over one electron Larmor radius. The weak-amplitude expansion, nonvanishing field, and weak variation over a gyro-orbit motivate the local-response approximation; they do not provide an error bound for the eigenfrequency or growth rate. The nominal, caution, marginal, and outside labels used below indicate the adopted ordering diagnostics, rather than validation against a full nonuniform kinetic theory.

There is also a specific geometric restriction in the scans. With \(\bm{B}=B(y)\hat{\bm{z}}\), the usual guiding-center magnetic-gradient drift has direction \(\bm{v}_{\nabla B}\propto\hat{\bm{b}}\times\nabla B\), where \(\hat{\bm{b}}=\bm{B}/B\), and hence lies along the \(x\) axis. Here the drift velocity is divided by \(c_s\), consistently with the normalized wave vectors and frequencies. All retained sidebands use \(k_x=0\), so their wave vectors \(\bm{k}_m=(0,q+mK_B,k_z)\) satisfy \(\omega_{\nabla B,m}=\bm{k}_m\cdot\bm{v}_{\nabla B}=0\). Thus, within this guiding-center description, the direct magnetic-drift contribution to the resonance frequency vanishes in the chosen geometry. This argument addresses that specific drift term only. It neither validates the full high-frequency susceptibility in a nonuniform field nor eliminates finite-orbit corrections, variation sampled along the \(E\times B\) motion, or equilibrium adjustments. In a geometry with a nonzero wave-vector component along the magnetic drift, the corresponding resonance would have to be retained.

The model also assumes prescribed background quantities such as \(\bm{E}_0\), density, and electron temperature. It does not solve a self-consistent nonuniform equilibrium. Its homogeneous-background susceptibility omits explicit gradient-driven terms associated with \(\nabla n\), \(\nabla T\), \(\nabla B\), and \(\nabla V_E\). This restriction follows from the adopted equilibrium and response operator, rather than from using fluid ions with kinetic electrons. In particular, inserting the prescribed \(B(y)\) dependence into the uniform-field response and coupling its Fourier sidebands does not recover the equilibrium-gradient terms that drive Simon--Hoh and related gradient-drift modes.\cite{Hara2022GradientDrift,Romadanov2016Nonlocal} A small \(\chi\) constrains magnetic variation on an electron-orbit scale; it does not establish that the omitted long-wavelength branches have negligible growth.

Including those branches would require a consistent equilibrium and a linear response retaining the relevant gradients and spatial structure, with appropriate boundary conditions where a nonlocal treatment is needed. A Bloch--Floquet expansion could still be applied if the resulting linear operator retained periodicity in the modulation direction, but the present matrix cannot be supplemented by those modes without rederiving its underlying response and couplings. Thus, the low-\(k_y\) results below describe only branches supported by the adopted kernel; no comparison of their growth with omitted gradient-drift modes is established here.

Finally, the model is linear: it predicts frequencies, growth rates, and sideband structure, but it does not determine nonlinear saturation or anomalous transport. A lower maximum linear growth rate, \(\gamma_{\max}(\epsB,\kb)<\gamma_{\max}(0)\), does not by itself imply lower electron mobility. Here \(\gamma_{\max}\) is the maximum growth rate over the wave-number interval being considered. Transport depends on fluctuation amplitudes, phases, nonlinear saturation, and correlations such as \(\langle \delta n_e\delta E_y\rangle\), where \(\delta n_e\) is the electron-density perturbation and \(\delta E_y\) is the azimuthal electric-field perturbation, as emphasized in instability-based transport theories.\cite{Lafleur2016Part1,Lafleur2016Part2}

Within these restrictions, the semi-local Floquet construction provides an approximate linear screening model for resonance detuning and sideband coupling. It identifies whether a prescribed magnetic modulation has a direct linear tendency to reduce the drift-instability growth spectrum, while separating that question from nonlinear saturation and transport closure. Candidates that reduce the full linear-growth envelope within the model ordering would merit further investigation for a direct linear stabilization mechanism. A lack of reduction lowers their priority on that basis, but cannot rule out beneficial effects involving nonlinear saturation or equilibrium changes. Theoretical screening therefore helps prioritize subsequent simulations and experiments; it does not replace them or establish changes in electron mobility.

\section{Numerical solution strategy}
\label{sec:numerical_method}

\subsection{Scalar root extraction}
\label{sec:scalar_root_extraction}

Following uniform-field kinetic EDI formulations\cite{Ducrocq2006,Lafleur2016Part1,Lafleur2016Part2} and the fixed-point numerical implementation benchmarked by Cavalier \textit{et al.},\cite{Cavalier2013} the scalar calculation starts from the kinetic residual of Eq.~\eqref{eq:scalar_kinetic_residual}. Every Floquet sideband uses this response as its local kernel. For a fixed real wave vector \(\bm{k}=(k_x,k_y,k_z)\), define the electron response factor
\begin{equation}
  R_e(\omega,\bm{k};B_0)
  =
  \frac{k^2}{1+k^2+g(\Omega,X,Y)},
  \label{eq:electron_response_factor}
\end{equation}
where \(R_e\) is the scalar kinetic response appearing on the right-hand side of Eq.~\eqref{eq:scalar_kinetic_residual}. The scalar residual is then
\begin{equation}
  D_0(\omega,\bm{k};B_0)
  =
  (\omega-k_xV_p)^2
  -
  R_e(\omega,\bm{k};B_0).
  \label{eq:scalar_residual_method}
\end{equation}
Here \(\bm{k}\) is the wave vector, \(R_e\) contains the Gordeev function evaluated at the arguments in Eq.~\eqref{eq:gordeev_args}, and \(D_0=0\) is equivalent to the uniform dispersion relation. Numerically, the infinite sum in Eq.~\eqref{eq:gordeev_series} is replaced by a symmetric finite sum,
\begin{equation}
  g_{N_g}(\Omega,X,Y)
  =
  \frac{\Omega}{\sqrt{2Y}}
  \sum_{p=-N_g}^{N_g}
  \Gamma_p(X)
  Z\!\left(
  \frac{\Omega-p}{\sqrt{2Y}}
  \right),
  \label{eq:gordeev_truncated}
\end{equation}
where \(N_g\) is the retained harmonic half-width for the Gordeev evaluation. In one evaluation of \(D_0\), \(N_g\) is fixed so that the residual is a well-defined complex function of \(\omega\). For the parameter scans reported in Sec.~\ref{sec:numerical_results}, the default value is chosen from the local finite-Larmor-radius argument as
\begin{equation}
  N_g(X)
  =
  \max\left[
    24,\,
    \left\lceil 12\sqrt{X+1}+20\right\rceil
  \right].
  \label{eq:gordeev_count_rule}
\end{equation}
This empirical rule is conservative: it keeps many more cyclotron harmonics than the peak width of \(\Gamma_p(X)\) and was checked against larger explicit harmonic counts in scalar uniform-field root reproduction tests. If a specific convergence test is being performed, \(N_g\) can be overridden by an explicit harmonic half-width.

Following the fixed-point construction used by Cavalier \textit{et al.}, the scalar solver freezes the Gordeev response at the previous iterate \(\omega^{(j)}\), where \(j\) is the iteration index, and solves the resulting quadratic equation analytically:
\begin{equation}
  \omega_{\pm}^{(j+1)}
  =
  k_xV_p
  \pm
  \left[
  R_e(\omega^{(j)},\bm{k};B_0)
  \right]^{1/2}.
  \label{eq:scalar_fixed_point_update}
\end{equation}
The signs \(+\) and \(-\) label the two square-root branches. In the magnetic-modulation scans, both square-root branches provide scalar sideband roots of \(D_0(\omega,k_{ym};B_0)=0\), including positive and negative real frequencies at each \(q\). These frequencies are used as initial guesses, called predictors, for the full Floquet matrix solve; they are not automatically roots of the coupled problem. Restricting the predictors to the \(+\) branch is not sufficient to recover the full unstable spectrum: the \(-\) branch is not discarded. For the real sinusoidal profile and the scan parameters \(k_xV_p=0\), reflection relates \((\omega,q)\) to \((-\omega^*,-q)\); it does not generally identify the two branches at the same \(q\). Both predictor branches, followed by matrix correction and independent root counts, are therefore included when constructing the growth envelope. These scalar predictors are then corrected at the Floquet determinant level, as described in Sec.~\ref{sec:floquet_root_method}.

The fixed-point scalar root is accepted only if it also satisfies a scale-free residual check,
\begin{equation}
  \eta_0
  =
  \frac{|D_0|}
  {\max\left(|(\omega-k_xV_p)^2|,\ |R_e|,\ 10^{-30}\right)}.
  \label{eq:scalar_quality}
\end{equation}
Equation~\eqref{eq:scalar_quality} should be read as a relative error for the scalar dispersion relation. The numerator \(|D_0|\) is the absolute mismatch between the cold-ion inertia term \((\omega-k_xV_p)^2\) and the electron response \(R_e\). The denominator compares this mismatch with the larger of the two terms being balanced; the \(10^{-30}\) floor only prevents division by a vanishing scale. The implementation stops when either \(\eta_0\) is below the fixed-point tolerance or the relative change in \(\omega\) is small and \(\eta_0\le\max(10^{-6},10\,{\rm tol})\). The default scalar fixed-point tolerance is \(10^{-9}\) for sideband predictors, and the maximum number of iterations is 80. The reported scalar growth rate is \(\gamma=\Im\omega\), and only roots with \(\gamma>0\) contribute to unstable-growth metrics.

\subsection{Ion azimuthal drift as a relative-drift test}
\label{sec:ion_rotation_method}

Before introducing magnetic modulation, it is useful to note a simpler way in which the same scalar dispersion relation can be modified. Zhou \textit{et al.}\ considered ion rotational flow in Hall-thruster azimuthal instability simulations and used the corresponding Doppler-shifted ion response as a theoretical guide.\cite{Zhou2025PLA} In the scalar normalized notation used here, a uniform ion azimuthal velocity \(V_i\equiv V_{i\theta}\) modifies the cold-ion inertia as
\begin{equation}
  D_i(\omega,\bm{k};V_i)
  =
  (\omega-k_xV_p-k_yV_i)^2
  -
  R_e(\omega,\bm{k};B_0)
  =
  0.
  \label{eq:ion_rotation_residual}
\end{equation}
The imposed ion azimuthal drift changes the ion inertial factor to \((\omega-k_xV_p-k_yV_i)^2\), while the laboratory-frame electron response \(R_e\) is unchanged. If the normalized frequency in the frame moving azimuthally with the ions is denoted by
\begin{equation*}
  \Omega_i=\omega-k_yV_i,
\end{equation*}
then the electron resonance argument in Eq.~\eqref{eq:gordeev_args} can be rewritten as
\begin{equation}
  \Omega
  =
  \frac{\omega-k_yV_d}{\omegace}
  =
  \frac{\Omega_i-k_y(V_d-V_i)}{\omegace}.
  \label{eq:relative_drift_argument}
\end{equation}
This form shows why the scalar growth is controlled primarily by the electron-ion azimuthal relative drift
\begin{equation*}
  V_{\rm rel}=V_d-V_i.
\end{equation*}
Here \(V_{\rm rel}\) is positive when the electron \(E\times B\) drift exceeds the co-directional ion azimuthal drift. Defining \(r=V_i/V_d\), co-drifting ions have \(r>0\) and reduce \(V_{\rm rel}/V_d=1-r\), whereas counter-drifting ions have \(r<0\) and increase the relative drift in this scalar linear model.

\begin{figure}[t]
\centering
\includegraphics[width=\linewidth]{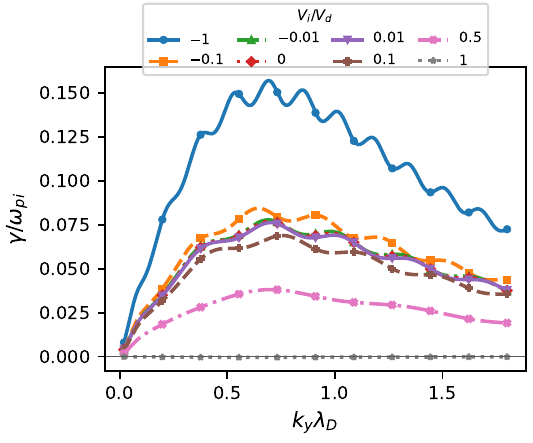}
\caption{Scalar kinetic growth spectra with imposed uniform ion azimuthal drift. The original axis labels \(k_y\lambda_D\) and \(\gamma/\omega_{pi}\) mean \(k_y^{\rm phys}\lambda_D=k_y\) and \(\gamma^{\rm phys}/\omega_{pi}=\gamma\) in the present notation. Co-drifting ions reduce the electron-ion relative drift and suppress the growth rate in the scalar model, but a substantial effect requires \(V_i\) to be comparable to the electron drift \(V_d\).}
\label{fig:ion_rotation_motivation}
\end{figure}

The numerical scan used for this relative-drift test is fully specified in the dimensionless normalization: frequencies are normalized by \(\omega_{pi}\), wave numbers by \(\lambda_D^{-1}\), and velocities by \(c_s\). Unless otherwise stated, the fixed parameters are
\begin{equation}
  k_x=0,\qquad
  k_z=0.045,\qquad
  V_p=0,
  \label{eq:rotation_scan_wavevector}
\end{equation}
and
\begin{equation}
  \begin{aligned}
    V_d&=150,\qquad v_{the}=491,\\
    \omegace&=50,\qquad M=2.4\times10^5 .
  \end{aligned}
  \label{eq:rotation_scan_params}
\end{equation}
These are the reference values taken from Table II of Cavalier \textit{et al.} and used throughout the scalar calculations, except that the axial ion drift \(V_p\) is set to zero in this diagnostic scan so that the only imposed ion drift is the azimuthal velocity \(V_i\). No separate dimensional density, temperature, or magnetic-field value is required to reproduce this normalized scan; the roots can be dimensionalized using a consistent reference state with \(c_s=\lambda_D\omega_{pi}\). The tabulated inputs \(M=2.4\times10^5\) and \(v_{the}=491\) retain their independently rounded values. The kinetic arguments use \(M\), corresponding to the consistent thermal speed \(\sqrt{M}=489.898\); 491 is used only for the ordering estimate \(\rho_e=491/50=9.82\). The resulting radius differs by approximately \(0.225\%\) from \(\sqrt{M}/50\). This rounding does not enter the dispersion residual and does not change any of the ordering classifications reported below. The scanned azimuthal wave-number interval is \(0.02\le k_y\le1.80\), sampled with 321 uniformly spaced points. The rotation ratio \(r=V_i/V_d\) is scanned over
\begin{equation*}
\begin{split}
  r\in\{&-1,-0.5,-0.1,-0.03,-0.01,-0.003,-0.001,0,\\
        &0.001,0.003,0.01,0.03,0.1,0.25,0.5,0.75,0.9,1\}.
\end{split}
\end{equation*}
Only a representative subset of these spectra is plotted in Fig.~\ref{fig:ion_rotation_motivation}. For each \(r\), the \(+\) square-root branch is followed by continuation in \(k_y\), starting from \(\omega=0\) at the first grid point. The rotating-ion diagnostic first applies the fixed-point iteration used by Cavalier \textit{et al.}\ and then corrects the resulting complex frequency with \texttt{scipy.optimize.root} using \texttt{method="hybr"}, the modified Powell hybrid method in MINPACK, applied to \(\Re D_i=0\) and \(\Im D_i=0\) as two real equations with a numerically estimated Jacobian. Roots are retained only when the normalized residual is below \(10^{-8}\). The fixed-point tolerance is \(10^{-10}\), and the maximum number of fixed-point iterations is 100. The Gordeev harmonic sum uses the same default rule in Eq.~\eqref{eq:gordeev_count_rule}, unless an explicit harmonic half-width is supplied for convergence testing.

Figure~\ref{fig:ion_rotation_motivation} shows the corresponding scalar growth spectra. The effect is strong only when \(V_i\) becomes a sizable fraction of \(V_d\). In the reference scan, a \(0.1\%\) co-drifting ion speed changes the maximum growth rate by only about \(0.1\%\), a \(1\%\) co-drifting ion speed changes it by about \(1\%\), and roughly \(10\%\) suppression requires \(V_i/V_d\simeq0.095\). With the order-of-magnitude values discussed by Zhou \textit{et al.}, \(V_d^{\rm phys}\sim5\times10^5\,{\rm m/s}\), so this \(10\%\) linear-growth reduction would require an ion azimuthal speed of order \(4.7\times10^4\,{\rm m/s}\), far above ordinary neutral-vortex or ion-rotation speeds. Direct ion rotation is therefore not a practical engineering actuator in this scalar estimate. Here it serves a limited role: it demonstrates that changing the parameters that enter the dispersion relation can strongly reduce the linear growth rate, motivating a systematic search for more practical parameter modulations, such as prescribed magnetic modulation.

\subsection{Floquet matrix roots and accuracy checks}
\label{sec:floquet_root_method}

The period and amplitude scans in this subsection use the reference \(k_z=0.045\). The additional parallel-wave-number scan in Sec.~\ref{sec:kz_scan} uses its own truncations, grid, and envelope checks, specified there. The Floquet problem replaces the scalar condition \(D_0=0\) by the matrix condition in Eq.~\eqref{eq:det_condition}. For a prescribed modulation wave number \(K_B\), modulation amplitude \(\epsB\), and quasi-wavenumber \(q\), the retained sidebands are \(k_{ym}=q+mK_B\) with \(m=-\orderN,\ldots,\orderN\). The matrix dimension is \(n_F=2\orderN+1\). For a target normalized wave-number interval \(k_y\in[k_{y,\min},k_{y,\max}]\), the reported scans use
\begin{equation}
  \begin{aligned}
    \orderN
    &\ge
    \max\left[
      N_{\min},\,
      \left\lceil
        \frac{k_{\rm edge}}{K_B}-\frac{1}{2}
      \right\rceil
    \right],
    \\
    k_{\rm edge}
    &=
    \max(|k_{y,\min}|,|k_{y,\max}|).
  \end{aligned}
  \label{eq:sideband_count_rule}
\end{equation}
where \(k_{\rm edge}\) is the largest magnitude of the target wave-number window. In the parameter scans, \(N_{\min}=3\) and \(k_{\rm edge}=1.20\). This lower bound is the minimum index needed for a sideband to reach \(k_{\rm edge}\) somewhere in the Bloch zone; it does not guarantee coverage of both signed edges at every \(q\), or convergence. The production truncations are selected using the additional checks below. The scans use \(\orderN=7,21,33,66,92\) for \(L_B=1,4,10,20,40~\mathrm{mm}\), respectively. The 20 mm value includes additional sidebands beyond the coverage minimum of 46, because projected high-wave-number modes can be sensitive to the truncation boundary even when the upper growth envelope is converged. The sampled grid is
\[
  q/K_B\in[-1/2,1/2],
\]
including both endpoints for plotting and trapezoidal integration; the two endpoints are physically equivalent in the infinite Floquet problem. The number of grid points is chosen as
\[
  N_q=\max\left[41,\left\lceil K_B/0.01\right\rceil+1\right].
\]

At each \(q\), the matrix \(\DF(\omega,q)\) is assembled from the scalar residual and its magnetic derivative. In the implementation, the derivative with respect to the dimensionless field scale \(s=B/B_0\) is evaluated by a centered finite difference,
\begin{equation}
  \begin{aligned}
    \left.
    \partial_s D_0(\omega,k_{ym};s)
    \right|_{s=1}
    &\simeq
    \frac{D_0^+-D_0^-}{2h_s},
    \\
    D_0^\pm
    &=
    D_0(\omega,k_{ym};1\pm h_s).
  \end{aligned}
  \label{eq:finite_difference_b}
\end{equation}
where \(h_s\) is the finite-difference step and \(D_0^\pm\) are the scalar residuals evaluated at \(s=1\pm h_s\). In the fixed-electric-field model, \(s\) simultaneously changes \(\omegace\) and \(V_d\) according to the scaling stated below Eq.~\eqref{eq:s_expansion}. The reported scans use \(h_s=10^{-4}\), chosen small enough to approximate the local derivative but not so small that roundoff dominates.

The determinant is evaluated through a tridiagonal factorization using its phase and logarithmic magnitude to avoid overflow or underflow. The corrected scans use a custom complex Newton solver for the tridiagonal determinant, evaluated through its analytic logarithmic derivative,
\begin{equation}
  \omega^{(j+1)}=\omega^{(j)}-
  \left[\partial_\omega\log\det\DF(\omega^{(j)},q)\right]^{-1},
  \label{eq:floquet_root_vector}
\end{equation}
with the step magnitude capped at 0.15 in normalized frequency units. The derivative with respect to frequency is analytic, using the plasma-dispersion-function identity \(Z'(z)=-2[1+zZ(z)]\); the magnetic derivative retains Eq.~\eqref{eq:finite_difference_b}. An initial guess is a complex frequency at which the matrix-root iteration starts; it is not a separate numerical algorithm or an accepted root. We collect these guesses from both scalar-frequency branches, already accepted roots at neighboring \(q\), symmetry-related roots, and continuation from a nearby modulation amplitude. Different sources can supply nearly identical guesses, and distinct guesses can converge to the same matrix root. Repeated guesses may be skipped, whereas converged frequencies are separately merged using the root-separation tolerance below. Every retained frequency must satisfy the matrix acceptance checks. Finally, we compare the number of distinct accepted roots with an independent contour count; if roots are missing, extra complex-frequency guesses and recursive contour subdivision extend the search. During deflated searches, the logarithmic derivative is replaced by \(\partial_\omega\log\det\DF-\sum_r(\omega-\omega_r)^{-1}\), where \(\omega_r\) are already accepted distinct roots; every new candidate is validated against the original, undeflated matrix. For the real sinusoidal modulation, reflection maps \((\omega,q)\) to \((-\omega^*,-q)\); the upper envelope must therefore be even in \(q\) for a symmetric sideband truncation.

Root quality is checked with determinant residuals, singular values, branch continuity, and truncation convergence. Let the singular values of \(\DF\) be \(\sigma_1\ge\cdots\ge\sigma_{n_F}\). The diagnostic used below is the smallest singular value
\begin{equation}
  \sigma_{\min}(\omega,q)
  =
  \sigma_{n_F}[\DF(\omega,q)],
  \label{eq:singular_value_diagnostic}
\end{equation}
and the corresponding right singular vector \(\phivec_\sigma\) is normalized by \(\|\phivec_\sigma\|_2=1\). A correction uses at most 45 iterations, with a final frequency-step tolerance of \(3\times10^{-11}\). All accepted frequencies and matrices are finite and satisfy \(\sigma_{\min}<10^{-8}\). A transferred root at an exactly singular factorization pivot is retained directly only if \(\sigma_{\min}<10^{-11}\). Distinct roots are separated with a complex-frequency tolerance of \(2\times10^{-7}\). The unstable-root audit uses \(\gamma>10^{-7}\); this cutoff lies far below every reported upper envelope. For an accepted root, the normalized sideband weights are
\begin{equation}
  \begin{gathered}
  w_m=\frac{|\phi_m|^2}{\displaystyle\sum_{j=-\orderN}^{\orderN}|\phi_j|^2},\\
  m=-\orderN,\ldots,\orderN,\qquad \sum_{m=-\orderN}^{\orderN}w_m=1.
  \end{gathered}
  \label{eq:sideband_weights}
\end{equation}
These weights are independent of the overall eigenvector normalization and identify the dominant projected wave number and the participation ratio.

The number of accepted unstable roots is checked against adaptive contour winding counts in two normalized-frequency rectangles: \(-1.6<\Re\omega<1.6\), \(10^{-7}<\Im\omega<0.6\), and \(-3<\Re\omega<3\), \(10^{-7}<\Im\omega<1.2\). Scalar-denominator poles are counted separately; none occurs within these audited rectangles. Counts agree at all 1308 parameter--grid combinations. At the zone edge and the grid point closest to zero, 60 additional checks over all parameter pairs expand the rectangle to \(-10<\Re\omega<10\), \(10^{-7}<\Im\omega<3\), without changing the counts. These are finite-domain checks, rather than a proof about the entire complex-frequency plane. Independent evaluations with the original matrix implementation give a largest selected-root singular-value residual of \(3.60\times10^{-13}\).

All absolute frequency and envelope differences reported here are dimensionless, i.e., their dimensional values divided by \(\omega_{pi}\). At zero modulation, independent solutions of both scalar-frequency branches and fresh matrix corrections agree for all five periods at \(q/K_B=-0.5,-0.25,0\). Across these 15 tests, all 1324 distinct unstable roots match in the larger finite rectangle stated above, with a maximum nearest-root frequency-set difference of \(2.42\times10^{-16}\); the off-diagonal entries vanish exactly. This numerically verifies recovery of the folded uniform-field unstable spectrum in the tested domain.

\begin{figure*}[tbp]
  \centering
  \includegraphics[width=0.86\textwidth]{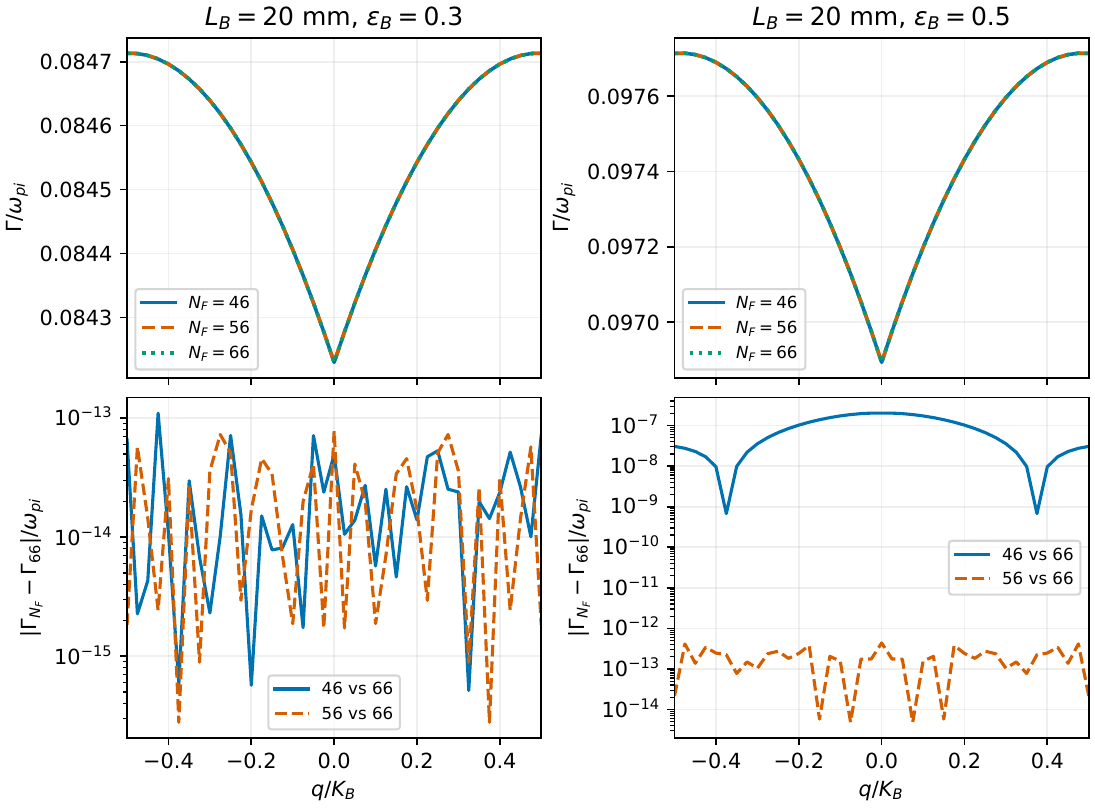}
  \caption{Sideband-truncation convergence for \(L_B=20~\mathrm{mm}\) at \(\epsB=0.3\) (left) and 0.5 (right), using \(h_s=10^{-4}\) and the same 41 Bloch points. Upper panels compare the growth-rate envelopes for \(\orderN=46,56,66\); the retained ordinate \(\Gamma/\omega_{pi}\) means \(\Gamma^{\rm phys}/\omega_{pi}=\Gamma\). Lower panels show absolute differences from \(\orderN=66\). Their retained axis labels mean \(|\Gamma_{N_F}^{\rm phys}-\Gamma_{66}^{\rm phys}|/\omega_{pi}=|\Gamma_{N_F}-\Gamma_{66}|\); dimensionless differences below \(10^{-16}\) are displayed at that floor only for logarithmic plotting. The nearly coincident envelopes establish weak truncation sensitivity of this diagnostic; projected-window convergence is checked separately in the text.}
  \label{fig:nf_convergence}
\end{figure*}

Figure~\ref{fig:nf_convergence} explicitly compares \(\orderN=46,56,66\) for \(L_B=20~\mathrm{mm}\) at \(\epsB=0.3,0.5\), with \(h_s=10^{-4}\) and the same 41 Bloch points. Changing \(\orderN=46\) to 66 changes \(\Gamma\) by at most \(2.02\times10^{-7}\), or a relative \(2.07\times10^{-6}\) of the corresponding reference peak. Envelope convergence alone does not establish convergence of individual modes or projected windows. The 20 mm results use the padded \(\orderN=66\); between 56 and 66, the projected-window peaks and integrals change by at most a relative \(4.98\times10^{-6}\) across all five windows and six amplitudes. The other periods are checked with two larger truncations by correcting leading envelope roots and additional outer-sideband scalar predictors; the largest production-to-largest-truncation envelope change is \(6.27\times10^{-12}\).

Table~\ref{tab:consistency_summary} summarizes independent step and grid tests for three representative modulated cases, using \(\orderN=66\) at 20 mm and 92 at 40 mm. The centered magnetic-derivative steps are \(h_s=2\times10^{-4},10^{-4},5\times10^{-5}\), at fixed \(N_q=41\); the nested Bloch grids have \(N_q=41,81,161\), at fixed \(h_s=10^{-4}\). Here \(\Gamma_{\max}=\max_q\Gamma(q)\), and \(I_q\) is the positive-envelope integral over \(q/K_B\in[-1/2,1/2]\), evaluated by the trapezoidal rule. Halving the production step changes \(I_q\) by at most a relative \(5.08\times10^{-7}\). Relative to 161 points, the largest \(I_q\) differences across the three modulated cases and their two uniform-field baselines are \(3.52\times10^{-6}\) for 41 points and \(7.04\times10^{-7}\) for 81 points. Both global growth metrics remain enhanced relative to the corresponding uniform-field baseline in all three modulated cases.

For the additional step and grid solves, recovered roots agree with contour counts in the upper-frequency strip \(-3<\Re\omega<3\), \(\gamma_{\rm cut}<\Im\omega<1.2\), where \(\gamma_{\rm cut}\) is 95\% of the minimum audited production envelope for that parameter pair. All reported envelope roots lie above this cutoff. There are 426 independently counted new half-grid points and 105 reused production points; reflection supplies positive \(q\), with original-matrix residual checks. These tests support the global upper-envelope metrics in the tested cases, not step/grid convergence of every individual mode or projected-window integral. Finite-domain and finite-grid agreement is not a proof over the entire frequency plane or continuous \(q\), and numerical convergence does not extend the physical validity of the semi-local model.

\begin{table*}[tbp]
  \caption{Step and Bloch-grid sensitivity of the global growth metrics. Step columns compare \(h_s=10^{-4}\) with \(5\times10^{-5}\) at \(N_q=41\); \(\Delta_h\Gamma\) is the maximum pointwise absolute difference of the dimensionless envelopes, equal to the dimensional difference divided by \(\omega_{pi}\). Grid columns compare \(N_q=41\) with 161 at \(h_s=10^{-4}\). For a positive metric \(\mathcal{M}\), \(\delta\mathcal{M}=|\mathcal{M}-\mathcal{M}_{\rm ref}|/\mathcal{M}_{\rm ref}\), with the smaller step or finer grid as reference. All \(\delta\) entries are dimensionless fractions, not percentages. \(\Gamma_{\max}\) is the envelope peak and \(I_q\) its positive Bloch-zone integral. The retained half-widths are \(\orderN=66\) at 20 mm and 92 at 40 mm.}
  \label{tab:consistency_summary}
  \centering
  \begin{tabular}{cccccc}
  \toprule
  \(L_B\) (mm) & \(\epsB\) & Step: \(\Delta_h\Gamma\) & Step: \(\delta I_q\) & Grid: \(\delta\Gamma_{\max}\) & Grid: \(\delta I_q\) \\
  \midrule
  20 & 0.3 & \(9.03\times10^{-9}\) & \(1.06\times10^{-7}\) & \(1.24\times10^{-6}\) & \(2.30\times10^{-6}\) \\
  20 & 0.5 & \(4.95\times10^{-8}\) & \(5.05\times10^{-7}\) & \(2.49\times10^{-6}\) & \(3.52\times10^{-6}\) \\
  40 & 0.5 & \(4.97\times10^{-8}\) & \(5.08\times10^{-7}\) & \(1.36\times10^{-6}\) & \(8.75\times10^{-7}\) \\
  \bottomrule
  \end{tabular}
\end{table*}

\section{Numerical results}
\label{sec:numerical_results}

\subsection{\texorpdfstring{Floquet sidebands and the \(q\)-space growth envelope}{Floquet sidebands and the q-space growth envelope}}
\label{subsec:q_space_envelope}

The numerical scans are interpreted in the Bloch representation rather than by assigning each computed root to a single azimuthal wave number. This distinction is essential for a periodically modulated magnetic field. For a fixed magnetic modulation wave number \(K_B\), a Floquet eigenmode at quasi-wavenumber \(q\) contains the sidebands \(k_{ym}=q+mK_B\), where \(m\) is an integer sideband index. These sidebands are not independent when \(\epsB\neq0\). They are coupled through the off-diagonal entries of \(\DF\), and a single matrix root may contain appreciable weight in more than one sideband. Therefore, the most direct stability question is not whether a selected projected \(k_y\) branch is locally reduced, but whether the most unstable root at a given Bloch quasi-wavenumber is reduced.

\begin{figure}[tp]
  \centering
  \includegraphics[width=0.35\textwidth]{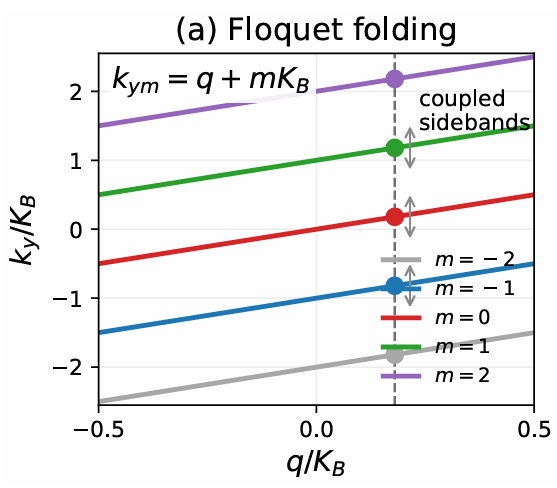}
  \includegraphics[width=0.35\textwidth]{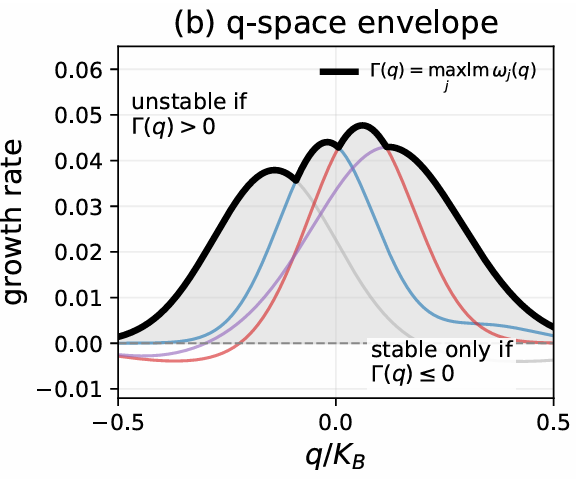}
  \caption{Schematic relation between Floquet sidebands and the \(q\)-space growth diagnostic. (a) In the first Brillouin zone, \(q/K_B\in[-1/2,1/2]\), a single Floquet eigenmode contains sidebands \(k_{ym}=q+mK_B\). The vertical dashed line indicates one fixed \(q\); the colored points are the coupled sidebands retained in the matrix problem. (b) At each \(q\), the matrix dispersion relation can return several roots \(\omega_j(q)\). The stability diagnostic used in this paper is the upper envelope \(\Gamma(q)=\max_j\operatorname{Im}\omega_j(q)\), not a single dominant-\(k_y\) projection. A true stable Bloch interval would require \(\Gamma(q)\le0\) over that interval. The plotted curves are schematic and are not numerical data.}
  \label{fig:floquet_sideband_q_envelope}
\end{figure}

Figure~\ref{fig:floquet_sideband_q_envelope} illustrates this diagnostic. The quasi-wavenumber \(q\) is restricted to the first Brillouin zone, while the physical azimuthal wave numbers are represented by the folded sideband set. This restriction loses no physical Fourier components. If \(q\) is shifted by an integer multiple of the modulation wave number, \(q'=q+nK_B\), then
\begin{equation}
  q'+mK_B=q+(m+n)K_B .
  \label{eq:q_relabeling}
\end{equation}
Thus the sideband set generated from \(q'\) is the same set generated from \(q\), with the sideband index relabeled from \(m\) to \(m+n\). Equivalently, Bloch quasi-wavenumbers separated by a reciprocal-lattice vector \(K_B\) describe the same physical packet of azimuthal Fourier components. One may therefore choose a single representative \(q\) from each equivalence class, conventionally \(-K_B/2<q\le K_B/2\). In the infinite-sideband problem this folding is exact; in a finite truncation it is recovered to the accuracy checked by increasing \(\orderN\). Let \(\omega_j(q)=\omega_{r,j}(q)+\ii\gamma_j(q)\) denote the \(j\)-th accepted Floquet root at the same \(q\), where \(\omega_{r,j}\) is its real frequency and \(\gamma_j\) is its growth rate. The envelope
\begin{equation}
  \Gamma(q)=\max_j \gamma_j(q)
  =
  \max_j \operatorname{Im}\omega_j(q)
  \label{eq:q_space_envelope}
\end{equation}
is then the relevant linear stability measure at that Bloch quasi-wavenumber. A modulation produces a genuine stable Bloch interval only if \(\Gamma(q)\le0\) over a finite interval of \(q\). For compact comparison between different modulation amplitudes, the calculations also report the maximum value \(\Gamma_{\max}=\max_q\Gamma(q)\) and the normalized integrated positive growth
\[
  I_q=\int_{-1/2}^{1/2}\max[\Gamma(\xi K_B),0]\,\dd \xi,
  \qquad \xi=q/K_B .
\]
This is the quantity computed by the numerical procedure using trapezoidal integration over the sampled coordinate \(q/K_B\). Both \(I_q\) and \(\Gamma\) are dimensionless. Integration over the normalized wave number \(q\) instead gives \(K_B I_q\). With \(\Gamma^{\rm phys}=\omega_{pi}\Gamma\) and \(q^{\rm phys}=q/\lambda_D\), the fully dimensional integral over one Bloch zone is
\[
  \int_{-K_B^{\rm phys}/2}^{K_B^{\rm phys}/2}
  \max[\Gamma^{\rm phys}(q^{\rm phys}),0]\,\dd q^{\rm phys}
  =\frac{\omega_{pi}}{\lambda_D}K_B I_q .
\]
These fixed conversion factors cancel in amplitude-suppression ratios at the same \(L_B\) and reference state.

The dominant-\(k_y\) projection remains useful, but only as an interpretive diagnostic. From the right singular vector associated with a root, the sideband weights \(w_m\) defined in Eq.~\eqref{eq:sideband_weights} identify a dominant sideband \(m_\ast\) and a projected wave number \(k_{y\ast}=q+m_\ast K_B\). This projection helps show whether the modulation redistributes spectral weight between MTSI-like and ECDI-like wave-number ranges. However, the projection can display apparent gaps, jumps, or local reductions when the dominant sideband changes. Such features do not by themselves demonstrate stabilization. The growth comparisons below are therefore based first on the \(q\)-space envelope in Eq.~\eqref{eq:q_space_envelope}; projected \(k_y\) plots are used only to explain how the spectrum is redistributed.

\subsection{\texorpdfstring{Long-wavelength amplitude scan at \(L_B=20~\mathrm{mm}\)}{Long-wavelength amplitude scan at LB=20 mm}}
\label{subsec:lb20_amplitude_scan}

The first numerical case is a long-wavelength modulation with imposed magnetic period \(L_B=20~\mathrm{mm}\). The dimensionless modulation wave number used in the Floquet calculation is
\begin{equation}
  K_B=\frac{2\pi\lambda_D}{L_B}=0.02617994,
  \qquad \lambda_D=\frac{1}{12}~\mathrm{mm},
  \label{eq:lb20_kb}
\end{equation}
where \(\lambda_D\) is the reference Debye length. The scan uses \(k_x=0\), \(k_z=0.045\), six amplitudes \(\epsB=0,0.1,0.2,0.3,0.4,0.5\), and 41 Bloch points over \(q/K_B\in[-1/2,1/2]\). The projected interval extends to the normalized wave number \(k_y=1.20\), whose coverage minimum from Eq.~\eqref{eq:sideband_count_rule} is \(\orderN=46\). The window and envelope results use \(\orderN=66\), with 133 retained sidebands, to separate the target interval from the truncation boundary. For the normalized radius \(\rho_e=9.82\), \(\chi=\epsB K_B\rho_e\) is nominally valid up to \(\epsB=0.3\); \(\epsB=0.4,0.5\) require caution.

\begin{figure*}[tbp]
  \centering
  \includegraphics[width=0.75\textwidth]{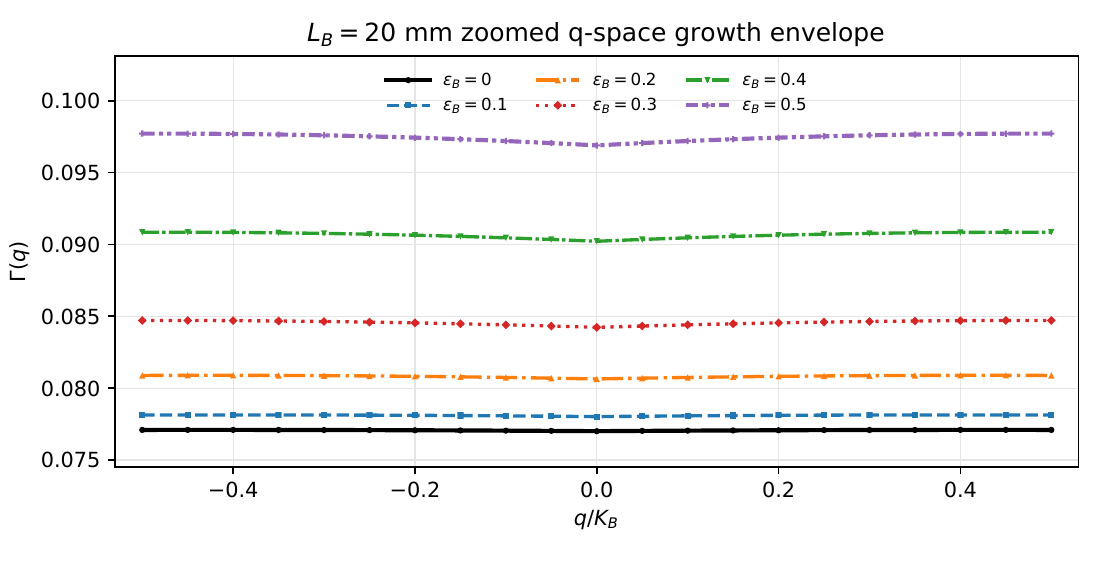}
  \caption{Audited \(q\)-space growth envelopes for \(L_B=20~\mathrm{mm}\), using \(\orderN=66\). The envelope \(\Gamma(q)=\max_j\operatorname{Im}\omega_j(q)\) includes both signs of real frequency. All sampled points remain unstable, and the nonzero-amplitude envelopes exceed the uniform reference.}
  \label{fig:lb20_q_envelope_overlay}
\end{figure*}

Figure~\ref{fig:lb20_q_envelope_overlay} shows the audited upper envelopes. The uniform-field reference is nearly flat, with \(\Gamma\simeq0.077>0\), and is already unstable. At fixed \(L_B=20~\mathrm{mm}\), mean field \(B_0\), and remaining parameters, increasing \(\epsB\) increases the amplitude of magnetic nonuniformity. At every sampled \(q\), the upper growth envelope increases monotonically across the tested amplitudes \(\epsB=0,0.1,\ldots,0.5\). Including all counted unstable branches removes the apparent downward segments caused by incomplete root searches. The envelopes are even in \(q\), as required by reflection symmetry. The values in Table~\ref{tab:lb20_q_metrics} use the same root-coverage procedure for both the modulated cases and their uniform-field reference.

The largest-amplitude case, \(\epsB=0.5\), has \(\max_q\Gamma=0.09771\) and \(\min_q\Gamma=0.09689\). All sampled envelope values remain positive. The numerical increase is a property of the reduced Floquet matrix; the semi-local caution for the two largest amplitudes remains a separate physical limitation.

\begin{figure*}[tbp]
  \centering
  \includegraphics[width=0.9\textwidth]{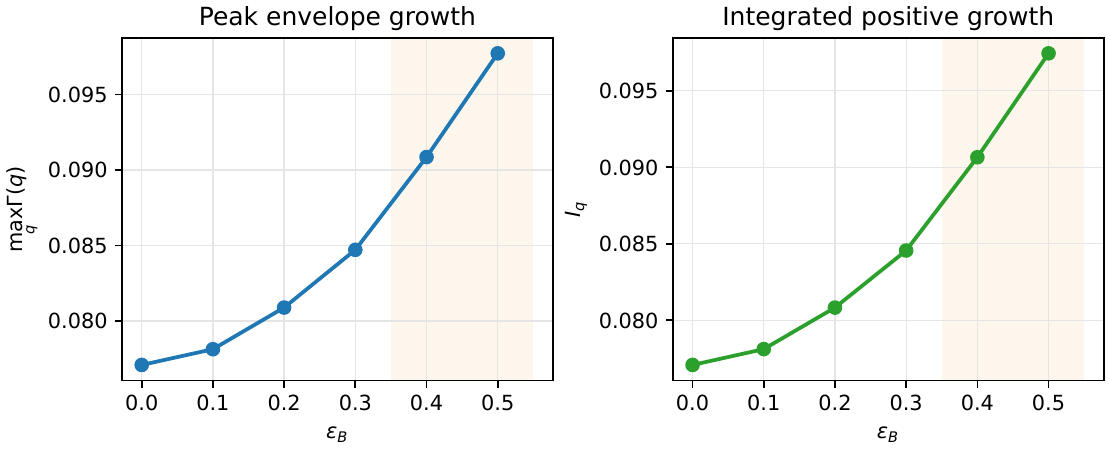}
  \caption{Amplitude dependence of the peak (left, blue) and integrated positive \(q\)-envelope growth (right, green) for \(L_B=20~\mathrm{mm}\). The corresponding suppression percentages are listed in Table~\ref{tab:lb20_q_metrics}. The shaded region marks the range \(\epsB\ge0.4\), which is treated with caution in the semi-local ordering.}
  \label{fig:lb20_amplitude_summary}
\end{figure*}

Table~\ref{tab:lb20_q_metrics} and Fig.~\ref{fig:lb20_amplitude_summary} summarize the global metrics. For any positive growth metric \(\mathcal{M}\), the suppression percentage is \(100[1-\mathcal{M}(\epsB)/\mathcal{M}(0)]\): positive values mean reduction and negative values mean increase. Both suppression percentages are negative for every nonzero tested amplitude, while the underlying peak and integral remain positive. At \(\epsB=0.5\), the peak increases by 26.74\% and the normalized positive Bloch-zone integral increases by 26.44\%. These increases indicate faster linear amplification of the unstable spectrum, rather than a transition from a stable reference state to instability. This scan therefore supplies no global linear-growth suppression within the tested parameter set.

\begin{table*}[tbp]
  \caption{Global \(q\)-envelope metrics for \(L_B=20~\mathrm{mm}\). The peak suppression is computed as \(100[1-\max_q\Gamma_{\epsB}(q)/\max_q\Gamma_{0}(q)]\), where \(\Gamma_0\) is the uniform-field envelope. Negative suppression means that the corresponding growth metric increases.}
  \label{tab:lb20_q_metrics}
  \begin{ruledtabular}
  \begin{tabular}{cccccccc}
  \(\epsB\) & \(\chi\) & validity & \(\max_q\Gamma\) &
  peak supp. & \(I_q\) & \(I_q\) supp. & \(\min_q\Gamma\) \\
   & & & & (\%) & & (\%) & \\
  \hline
  0.0 & 0.0000 & good & 0.07710 & 0.00 & 0.07707 & 0.00 & 0.07701 \\
  0.1 & 0.0257 & good & 0.07814 & -1.35 & 0.07811 & -1.34 & 0.07802 \\
  0.2 & 0.0514 & good & 0.08089 & -4.93 & 0.08082 & -4.87 & 0.08064 \\
  0.3 & 0.0771 & good & 0.08471 & -9.88 & 0.08455 & -9.71 & 0.08423 \\
  0.4 & 0.1028 & caution & 0.09085 & -17.84 & 0.09065 & -17.63 & 0.09022 \\
  0.5 & 0.1285 & caution & 0.09771 & -26.74 & 0.09745 & -26.44 & 0.09689
  \end{tabular}
  \end{ruledtabular}
\end{table*}

\rev{To describe redistribution in projected wave-number space, the accepted roots are also grouped into several projected-\(k_y\) windows. The grouping is made after assigning each root a dominant sideband wave number \(k_{y\ast}=q+m_\ast K_B\), as defined in Sec.~\ref{subsec:q_space_envelope}. All window limits below use the dimensionless wave number \(k_y\) defined in Eq.~\eqref{eq:normalized_variables}. The MTSI low-\(k_y\) curve contains roots with \(0.04\le k_{y\ast}\le0.25\), the ECDI \(n=1\) curve uses \(0.20\le k_{y\ast}\le0.50\) around the first electron-cyclotron resonance \(k_y\simeq\omega_{ce}/V_d=0.333\), the ECDI \(n=2\)/peak curve uses \(0.50\le k_{y\ast}\le0.95\) around \(2\omega_{ce}/V_d=0.667\) and the scalar peak near \(k_y\simeq0.70\), the high-\(k_y\) tail uses \(0.95\le k_{y\ast}\le1.20\), and the full curve uses \(0.04\le k_{y\ast}\le1.20\). These windows are diagnostic rather than a disjoint mode classification; in particular, the small overlap between the MTSI and ECDI \(n=1\) windows reflects the gradual transition between the low-\(k_y\) and first-resonance ranges. The low-\(k_y\) window includes only roots of the present kernel, not the gradient-drift modes omitted in Sec.~\ref{sec:validity}. Neither growth reduction in this window nor the absence of an assigned unstable root in a projected bin demonstrates stabilization of the long-wavelength spectrum of an inhomogeneous discharge.}

\begin{figure*}[tbp]
  \centering
  \includegraphics[width=0.75\textwidth]{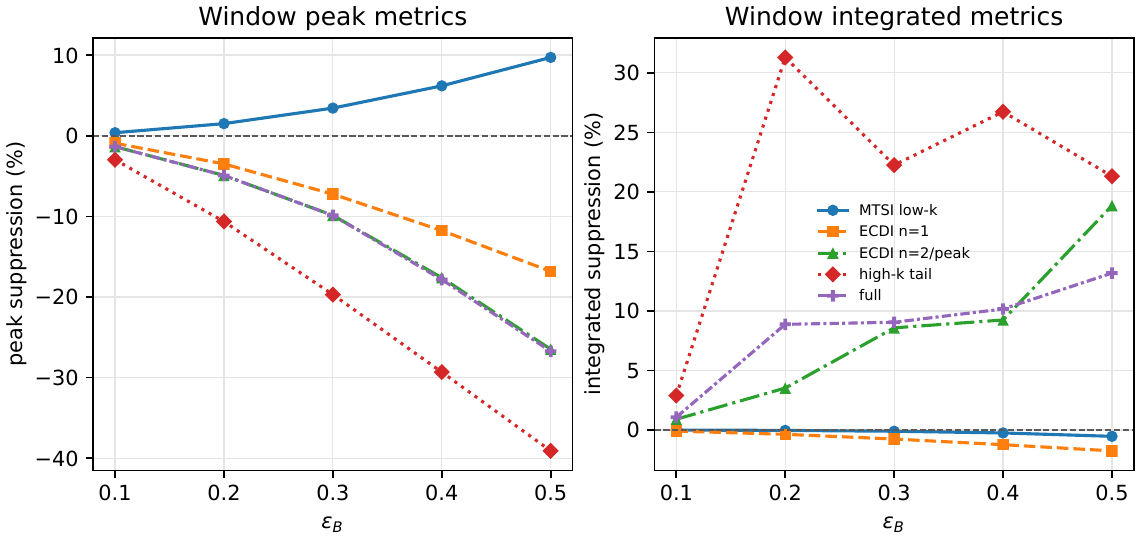}
  \caption{Peak and integrated suppression in projected-\(k_y\) windows: MTSI low-\(k_y\), 0.04--0.25; ECDI \(n=1\), 0.20--0.50; ECDI \(n=2\)/peak, 0.50--0.95; high-\(k_y\) tail, 0.95--1.20; and full interval, 0.04--1.20. Positive values denote reduction of the specified projected metric. Integrals include zero-valued unoccupied bins on the common projected grid; they are redistribution diagnostics, not evidence of a stable Bloch interval.}
  \label{fig:lb20_window_metrics}
\end{figure*}

Figure~\ref{fig:lb20_window_metrics} compares the five projected windows. At each grid value \(k_y=q+mK_B\), the projected positive-growth envelope is the maximum growth of roots assigned to that dominant sideband, or zero if no unstable root is assigned. Integrals use this common grid for every amplitude and the uniform reference; gaps are not bridged by integrating only the occupied bins. This definition makes the sampling explicit, but the projected integral remains sensitive to migration of the dominant-sideband label and is not a Bloch stability measure. Table~\ref{tab:lb20_window_eps05} lists the audited \(\epsB=0.5\) values. The MTSI-window peak is reduced by 9.71\%. The ECDI \(n=2\)/peak integrated measure is reduced by 18.85\%, and the high-\(k_y\) integral is reduced by 21.31\%. Any positive projected suppression must be interpreted alongside the increased full Bloch-zone envelope.

\begin{table}[tbp]
  \caption{Selected window metrics at \(L_B=20~\mathrm{mm}\) and \(\epsB=0.5\). Positive values indicate suppression relative to the uniform-field baseline.}
  \label{tab:lb20_window_eps05}
  \begin{ruledtabular}
  \begin{tabular}{lcc}
  window & peak supp. (\%) & integrated supp. (\%) \\
  \hline
  MTSI low \(k_y\) & 9.71 & -0.54 \\
  ECDI \(n=1\) & -16.82 & -1.76 \\
  ECDI \(n=2\)/peak & -26.49 & 18.85 \\
  high-\(k_y\) tail & -39.09 & 21.31 \\
  full interval & -26.74 & 13.18
  \end{tabular}
  \end{ruledtabular}
\end{table}

\begin{figure*}[tbp]
  \centering
  \includegraphics[width=0.8\textwidth]{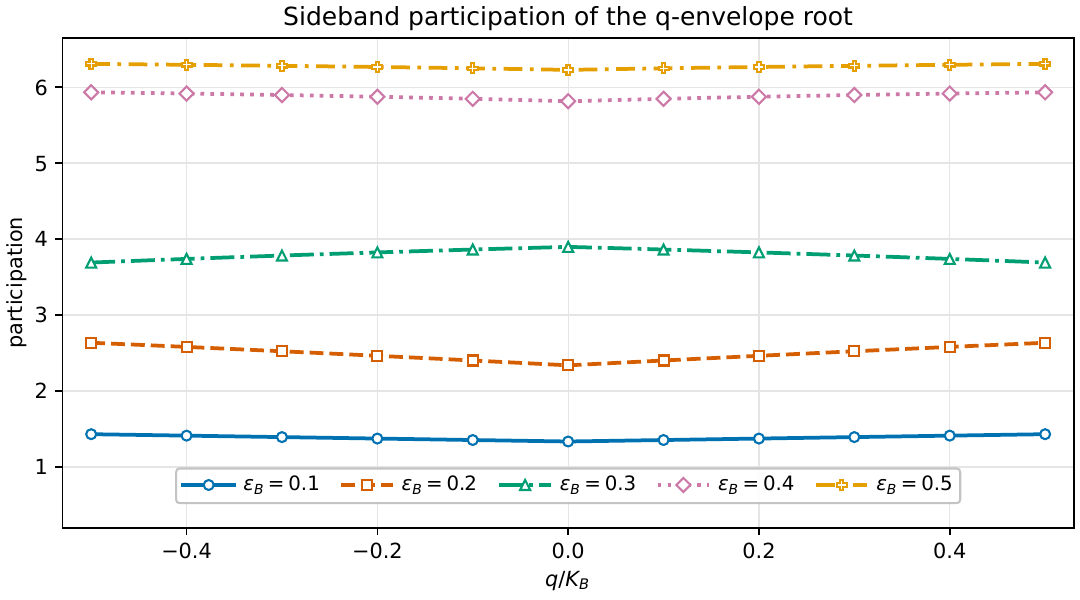}
  \caption{Sideband participation \(P(q)\) of the root that forms the \(q\)-space envelope. Larger values indicate that the envelope root contains several coupled Floquet sidebands rather than a single dominant scalar branch.}
  \label{fig:lb20_participation}
\end{figure*}

Sideband participation provides an additional mixing diagnostic. The participation ratio
\begin{equation}
  P(q)=\left(\sum_m w_m^2\right)^{-1}
  \label{eq:participation_ratio}
\end{equation}
measures how many sidebands contribute effectively to the root forming the envelope. Here \(w_m\) is the normalized sideband weight defined in Eq.~\eqref{eq:sideband_weights}. Values near unity indicate a root dominated by one sideband, while larger values indicate stronger Floquet mixing. Figure~\ref{fig:lb20_participation} shows that larger amplitudes produce stronger multi-sideband participation, supporting the conclusion that the long-wavelength modulation mainly redistributes spectral weight among coupled sidebands.

The 20 mm scan therefore shows persistent instability with increased global growth, together with reductions in selected projected-window measures. These statements are compatible because the window diagnostics and the Bloch-zone envelope quantify different aspects of the spectrum. A smaller MTSI-window peak does not imply a smaller maximum growth rate among all Floquet roots. Increased sideband participation likewise describes mode structure, rather than stabilization.

\subsection{\texorpdfstring{Global comparison over modulation wavelength and amplitude}{Global comparison over modulation wavelength and amplitude}}
\label{subsec:global_wavelength_comparison}

\begin{figure*}[tp]
  \centering
  \begin{minipage}{0.49\textwidth}
    \centering
    \includegraphics[width=\linewidth]{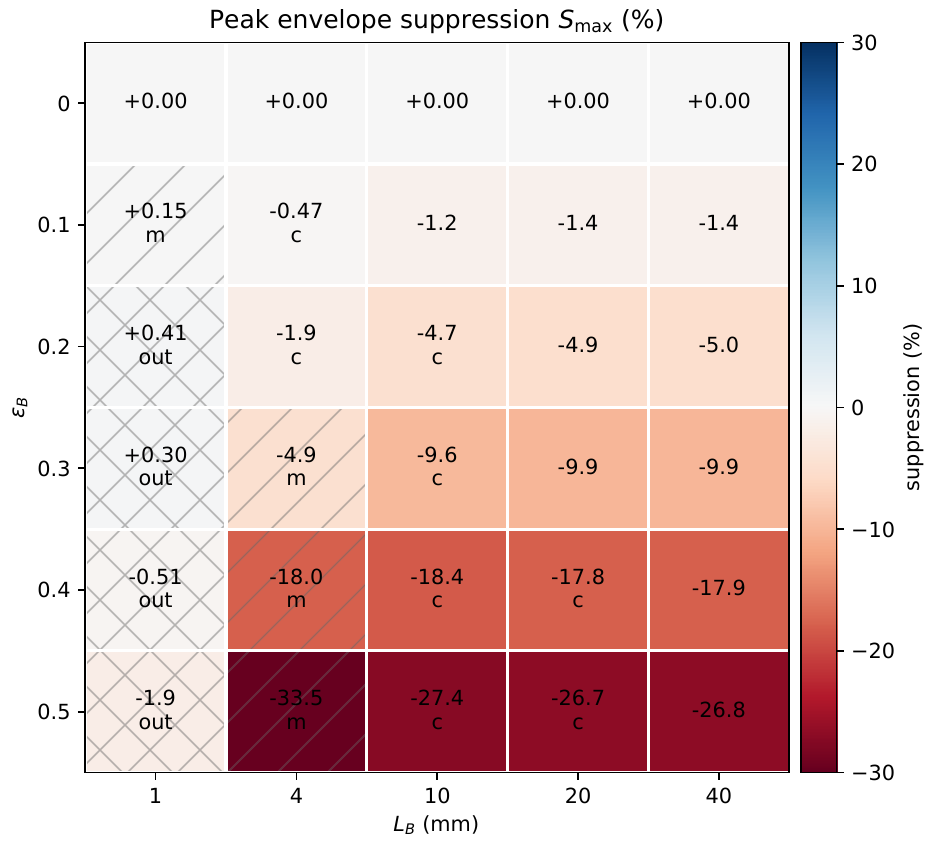}
  \end{minipage}
  \hfill
  \begin{minipage}{0.49\textwidth}
    \centering
    \includegraphics[width=\linewidth]{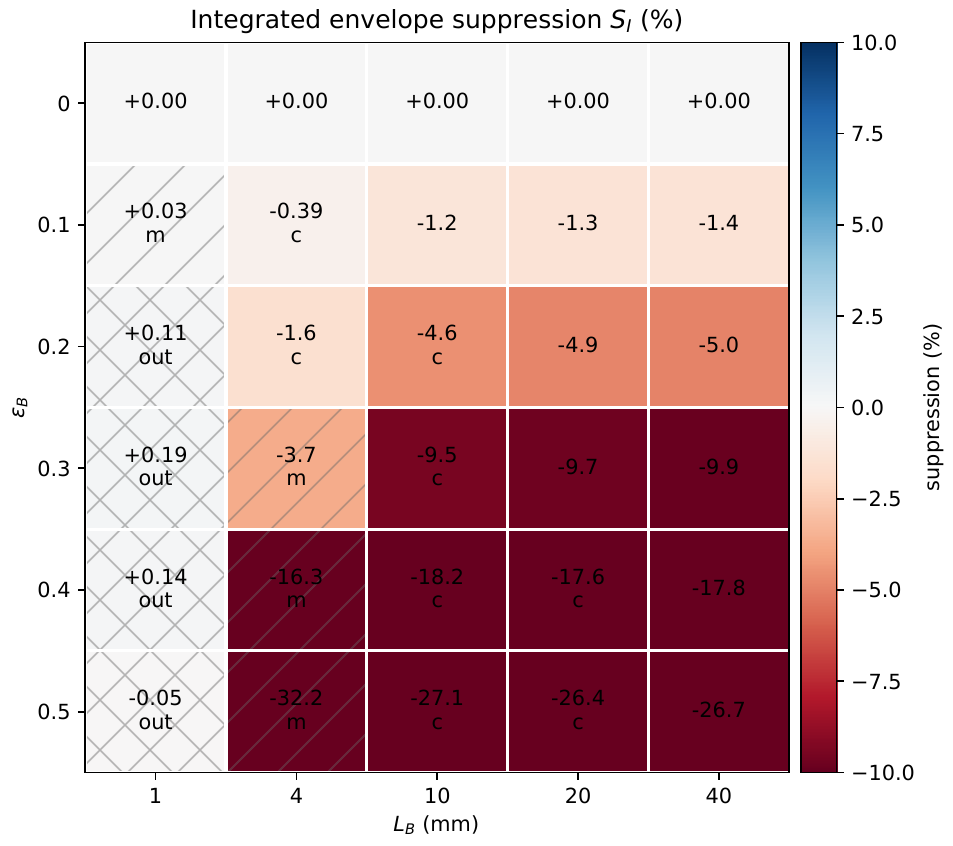}
  \end{minipage}
  \caption{Global suppression maps over modulation period \(L_B\) and magnetic-modulation amplitude \(\epsB\) on the common grid \(\epsB=0,0.1,0.2,0.3,0.4,0.5\). Left: peak \(q\)-envelope suppression \(S_{\max}\). Right: integrated positive \(q\)-envelope suppression \(S_I\). Blue positive values indicate lower growth metrics than the uniform-field baseline, while red negative values indicate higher growth metrics. The letters \(c\), \(m\), and \(out\) mark caution, marginal, and outside the semi-local validity ordering, respectively. The color scales use limits of \(\pm30\%\) for \(S_{\max}\) and \(\pm10\%\) for \(S_I\); colors saturate beyond these limits, and the cell annotations give the actual values.}
  \label{fig:global_suppression_maps}
\end{figure*}

The period scan compares the \(q\)-envelope metrics across several imposed magnetic periods,
\[
  L_B=1,\ 4,\ 10,\ 20,\ 40~\mathrm{mm},
\]
using the same definition of the Bloch-zone envelope in Eq.~\eqref{eq:q_space_envelope}. The corresponding modulation wave number is \(K_B=2\pi\lambda_D/L_B\), so shorter periods have larger \(\chi=\epsB K_B\rho_e\) at the same amplitude. This point is important when interpreting high-amplitude short-wavelength entries: they are useful as mathematical trends of the reduced Floquet matrix, but they provide weaker evidence for a physically valid semi-local prediction. Figure~\ref{fig:global_suppression_maps} uses the common amplitude grid \(\epsB=0,0.1,0.2,0.3,0.4,0.5\). The cell labels \(c\), \(m\), and \(out\) denote caution, marginal, and outside the semi-local ordering, respectively, while blank validity labels correspond to the nominal semi-local range.

Table~\ref{tab:chi_values} gives \(\chi\) for every test case in this scan. The values are calculated from the exact input expression
\begin{equation}
  \chi(\epsB,L_B)=\epsB\frac{2\pi(1/12~\mathrm{mm})}{L_B}\frac{491}{50}.
  \label{eq:chi_scan_values}
\end{equation}
The qualitative plotting labels use \(\chi<0.1\) (nominal), \(0.1\le\chi<0.3\) (caution, \(c\)), \(0.3\le\chi<1\) (marginal, \(m\)), and \(\chi\ge1\) (outside, \(out\)). These are diagnostic ranges for the necessary weak-gradient estimate, not rigorous error bounds or a sufficient validity test for the amplitude expansion.

\begin{table*}[tbp]
  \caption{Semi-local ordering parameter \(\chi\) for all wavelength--amplitude test cases in Sec.~\ref{subsec:global_wavelength_comparison}, rounded to six decimal places. Equation~\eqref{eq:chi_scan_values} specifies the unrounded values; the same reference \(\rho_e=9.82\) is used throughout.}
  \label{tab:chi_values}
  \centering
  \begin{tabular}{cccccc}
  \toprule
  \(\epsB\) & \(L_B=1\) mm & 4 mm & 10 mm & 20 mm & 40 mm \\
  \midrule
  0.0 & 0.000000 & 0.000000 & 0.000000 & 0.000000 & 0.000000 \\
  0.1 & 0.514174 & 0.128543 & 0.051417 & 0.025709 & 0.012854 \\
  0.2 & 1.028348 & 0.257087 & 0.102835 & 0.051417 & 0.025709 \\
  0.3 & 1.542522 & 0.385630 & 0.154252 & 0.077126 & 0.038563 \\
  0.4 & 2.056696 & 0.514174 & 0.205670 & 0.102835 & 0.051417 \\
  0.5 & 2.570870 & 0.642717 & 0.257087 & 0.128543 & 0.064272 \\
  \bottomrule
  \end{tabular}
\end{table*}

For each pair \((L_B,\epsB)\), the plotted peak metric is
\begin{equation}
  S_{\max}
  =
  100\left[
    1-\frac{\max_q\Gamma_{\epsB}(q)}
            {\max_q\Gamma_{0}(q)}
  \right],
  \label{eq:global_smax_metric}
\end{equation}
where \(\Gamma_0\) is obtained from the audited uniform-field roots at the same \(L_B\) and truncation. Positive \(S_{\max}\) means a lower peak. Figure~\ref{fig:global_suppression_maps} shows negative peak suppression for every tested nonzero amplitude at 10, 20, and 40 mm. The numerical increase is present both in the nominally valid subset and in the stronger-modulation cases marked for caution; their different levels of physical validity must still be respected. Small positive peak values at 1 mm occur only at the marginal or outside levels of the semi-local ordering and do not provide reliable evidence of physical stabilization. Thus the scan does not support a claim of global peak suppression in its nominally valid regime.

The integrated metric uses the normalized Bloch-zone integral defined in Sec.~\ref{subsec:q_space_envelope},
\begin{align}
  S_I
  =
  100\left[
    1-\frac{I_q(\epsB)}{I_q(0)}
  \right],
  \label{eq:global_integrated_metric}
  \\
  I_q(\epsB)=
  \int_{-1/2}^{1/2}
  \max[\Gamma_{\epsB}(\xi K_B),0]\,\dd\xi .
\end{align}
The right panel of Fig.~\ref{fig:global_suppression_maps} likewise shows increased integrated positive growth for every tested nonzero amplitude at 10, 20, and 40 mm. For 40 mm, \(S_I\) is \(-9.90\%\), \(-17.81\%\), \(-26.73\%\) at \(\epsB=0.3,0.4,0.5\), respectively. The small positive integrated entries at 1 mm occur only in marginal or outside cases. The complete counted branch sets therefore do not retain the apparent long-wavelength integrated reductions obtained from an incomplete root sample. Within the tested long-wavelength parameter set, the prescribed sinusoidal modulation redistributes and enhances the unstable spectrum. This is a statement about the computed linear growth metrics, not a general exclusion of suppression by other magnetic-field profiles or parameter choices.

\subsection{Sensitivity to the parallel wave number}
\label{sec:kz_scan}

We additionally vary the normalized parallel wave number over \(k_z=0.01,0.03,0.045,0.09\), with \(k_x=0\), \(L_B=20~\mathrm{mm}\), and \(\epsB=0,0.1,0.3\). All other inputs retain their reference values. Each modulated case is compared with the uniform-field envelope at the same \(k_z\), so a change in the baseline instability is not confused with the effect of modulation. The envelope uses 81 points across one Bloch zone, with \(N_F=106\) at \(k_z=0.01\) and \(N_F=66\) at the other three wave numbers. For the two nonzero amplitudes, \(\chi\) is 0.025709 and 0.077126, respectively, independent of \(k_z\); these are nominal weak-gradient ordering values, not bounds on the neglected nonuniform-orbit response.

The new calculations target the upper envelope. Independent contour counts locate and enumerate the roots in upper-frequency strips within \(-3<\Re\omega<3\), \(10^{-7}<\Im\omega<3\), discarding a higher strip only after its count is zero. For numerical conditioning, the contour determinant is divided by \((\omega+0.3\ii)^{2n_F}\), where \(n_F=2N_F+1\); this factor is analytic and nonzero in the upper half-plane and therefore does not change the winding count. Roots are corrected and checked using the original matrix. The negative half-zone is computed directly; the positive half follows reflection, with separate original-matrix checks for the reflected envelope roots. At the zone edge and center, upper-strip counts extended to \(|\Re\omega|<10\) find no higher growth rate. These checks concern the envelope in finite frequency domains, rather than a census of all weakly growing branches or a proof over the full complex plane. At \(k_z=0.01\) and \(\epsB=0.3\), a leading root localized at the edge of a smaller truncation did not persist after padding; the larger production set is retained on the basis of the following checks. At \(q/K_B=-0.5,-0.25,0\), we compare \(N_F=86,106,146\) for \(k_z=0.01\) and \(N_F=56,66,76\) for the other wave numbers, and halve the magnetic-derivative step. The largest relative envelope changes on increasing each production truncation to its largest tested value and on halving the step are \(7.94\times10^{-7}\) and \(1.32\times10^{-5}\), respectively. Refining the full-zone grid from 41 to 81 points changes the positive integral by at most \(2.47\times10^{-5}\) in relative terms.

\begin{figure*}[t]
  \centering
  \includegraphics[width=0.95\textwidth]{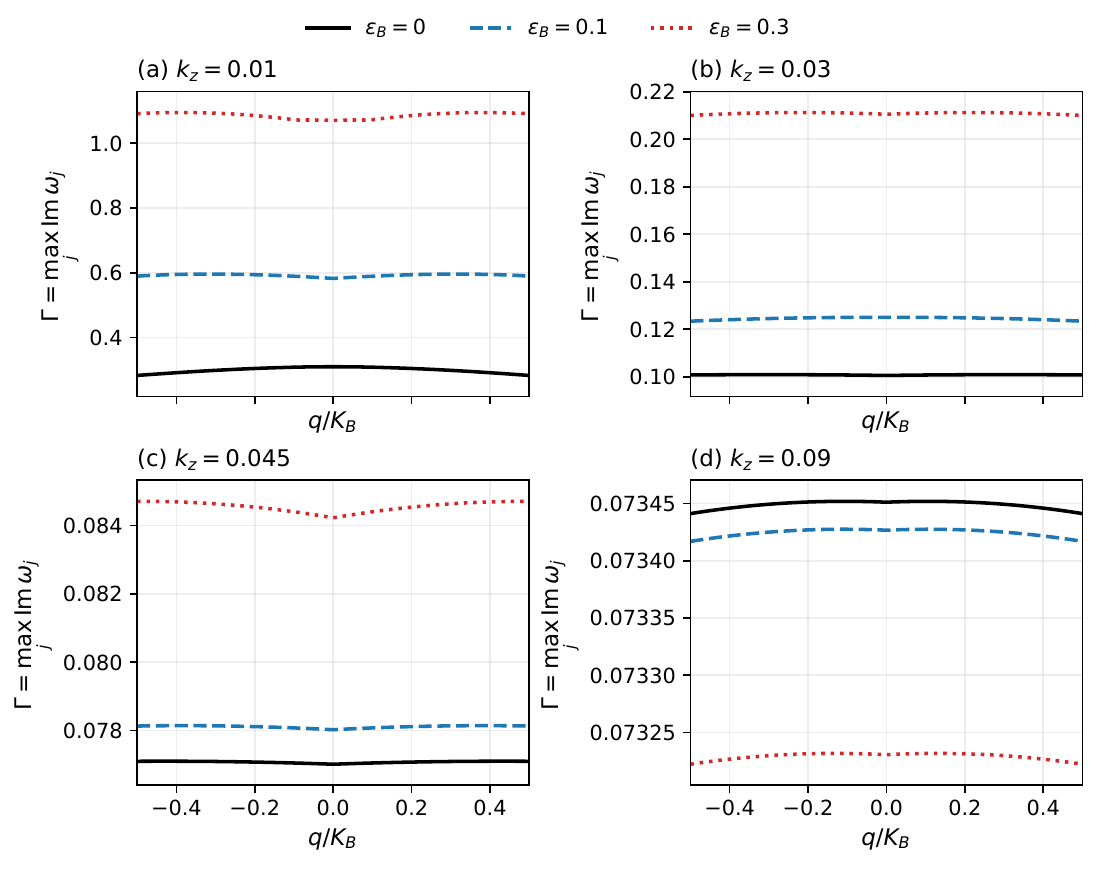}
  \caption{Parallel-wave-number scan at \(L_B=20~\mathrm{mm}\), with the normalized \(k_z\) indicated in each panel. Black solid, blue dashed, and red dotted curves denote \(\epsB=0,0.1,0.3\), preserving the amplitude styles used elsewhere. The ordinate is the dimensionless growth-rate envelope; panels use independent ordinate scales that do not start at zero, to resolve small changes. Curves contain 81 Bloch-zone points with \(N_F=106\) in panel (a) and \(N_F=66\) in panels (b)--(d). Each panel supplies its own uniform-field reference for the relative measures in Table~\ref{tab:kz_metrics}.}
  \label{fig:kz_envelopes}
\end{figure*}

Figure~\ref{fig:kz_envelopes} shows the envelopes and Table~\ref{tab:kz_metrics} gives the peak and integrated suppression measures, each relative to its own \(k_z\)-matched uniform reference. At \(k_z=0.01,0.03,0.045\), both measures are negative for the two tested nonzero amplitudes, indicating enhanced growth. At \(k_z=0.09\), however, the peak decreases by 0.033\% and 0.300\%, and the positive integral decreases by 0.033\% and 0.299\%, at \(\epsB=0.1\) and 0.3, respectively. All sampled envelopes remain positive: this is weak reduction of an already unstable spectrum, not stabilization. The largest tested \(k_z\) therefore gives the strongest reduction in this four-value sample, but the scan does not establish a continuous optimum. Numerical resolution of these small changes does not bound the error of the semi-local approximation or establish a measurable transport benefit.

Changing \(k_z\) changes the obliquity of each Fourier component. For \(k_x=0\), its angle to the magnetic field satisfies \(\tan\theta_m=|q+mK_B|/|k_z|\). A Floquet eigenmode contains multiple such components and therefore has no single propagation angle. The scan is accordingly reported in terms of \(k_z\), rather than assigning one angle to the entire coupled mode.

\begin{table}[htbp]
  \caption{Parallel-wave-number sensitivity of the global suppression measures (percent) at 20 mm. Positive entries mean reduction; negative entries mean enhanced growth relative to the uniform field at the same \(k_z\).}
  \label{tab:kz_metrics}
  \centering
  \begin{tabular}{ccccc}
  \hline
  & \multicolumn{2}{c}{\(\epsB=0.1\)} & \multicolumn{2}{c}{\(\epsB=0.3\)} \\
  \(k_z\) & \(S_{\max}\) & \(S_I\) & \(S_{\max}\) & \(S_I\) \\
  \hline
  0.01 & -92.279 & -97.299 & -253.106 & -261.556 \\
  0.03 & -23.909 & -23.522 & -109.354 & -109.195 \\
  0.045 & -1.351 & -1.344 & -9.881 & -9.712 \\
  0.09 & 0.033 & 0.033 & 0.300 & 0.299 \\
  \hline
  \end{tabular}
\end{table}

\section{Conclusions}
\label{sec:conclusions}

This work formulates a semi-local Floquet model for the linear effect of prescribed azimuthal magnetic-field modulation on high-frequency drift instabilities in Hall-thruster plasmas. The model uses a uniform-field kinetic EDI dispersion relation as a local spectral kernel and represents the periodic magnetic modulation through Floquet sideband coupling. In this construction, a scalar uniform-field condition \(D_0(\omega,k_y;B_0)=0\) is replaced by a matrix dispersion relation \(\det\DF(\omega,q;\epsB,\kb)=0\), where a Bloch quasi-wavenumber \(q\) labels a packet of coupled sidebands \(k_{ym}=q+mK_B\). The formulation is therefore a coupled-mode extension of the uniform kinetic response, not a complete kinetic theory with exact nonuniform electron orbits.

The numerical procedure combines both signs of scalar predictors, Floquet-root continuation, logarithmic-derivative correction, singular-value checks, contour root counts, and sideband-truncation tests. Stability is assessed from \(\Gamma(q)=\max_j\operatorname{Im}\omega_j(q)\), rather than an individual projected branch. Counting unstable roots in explicit finite frequency domains helps distinguish numerical branch loss from physical growth reduction. Projected wave-number windows and participation ratios remain useful structural diagnostics.

For \(L_B=20~\mathrm{mm}\) at the reference \(k_z=0.045\), the upper envelope increases with modulation amplitude at every sampled \(q\), and every nonzero tested amplitude raises both global envelope metrics. At \(\epsB=0.5\), the peak and integrated positive envelope increase by 26.74\% and 26.44\%, respectively. The uniform-field reference is already unstable, so the result represents faster linear amplification, not an instability threshold crossed from a stable state. The MTSI-window peak nevertheless decreases by 9.71\% in this case. Such a local projected reduction coexists with increased global growth and cannot establish global stabilization. The two largest amplitudes retain their semi-local caution labels.

The broader period scan at fixed \(k_z=0.045\), over \(L_B=1,4,10,20,40~\mathrm{mm}\), finds no global peak or integrated-growth reduction in its nominally valid nonzero-amplitude cases. At 10, 20, and 40 mm, both metrics increase for all tested nonzero amplitudes. The remaining small positive suppression values at 1 mm occur only in marginal or outside cases and cannot support a robust physical suppression claim. These results limit the control potential of the particular sinusoidal modulation within the tested parameters and the semi-local model. They do not imply that all magnetic modulations are destabilizing, and do not exclude suppression with different spatial profiles, other parameter regimes, or physics beyond this approximation. The root-coverage audit demonstrates why reliable branch enumeration is necessary before assigning physical significance to apparent envelope reductions.

These conclusions should be read within the assumptions of the semi-local model. The calculation assumes a prescribed magnetic modulation with nonzero field, a weak-to-moderate amplitude expansion, and a magnetic scale length that remains sufficiently long compared with the electron Larmor radius. It keeps the background density, temperature, and electric field prescribed; it does not include exact nonuniform electron-orbit effects, self-consistent equilibrium modification, explicit gradient-drift terms, nonlinear saturation, or the fluctuation correlations that determine anomalous electron transport. A response derived with the omitted gradient and orbit dynamics could change the calculated frequencies, growth rates, and stability boundaries; the magnitude and direction of those changes have not been established. The geometric cancellation of the direct magnetic-drift frequency for the selected scans is not a quantitative accuracy guarantee for the remaining response. Consequently, the observed increase in linear growth does not by itself establish a larger nonlinear fluctuation level or greater anomalous mobility. Likewise, a reduction of a projected linear-growth measure alone would not establish reduced transport.

The practical role of this framework is to guide the choice of magnetic profiles for subsequent simulations and experiments. The reference-wave-number period scan does not provide a promising direct linear stabilization case. The added \(k_z\) scan nevertheless shows that the sign can change: at \(k_z=0.09\), the 20 mm sinusoidal profile produces small reductions of the global growth measures. The strongest reduction among the sampled cases is 0.300\% in the peak and 0.299\% in the integral. The spectrum remains unstable, and the physical significance of such small changes requires validation beyond the semi-local approximation. These outcomes help distinguish parameter-sensitive tendencies before a broader simulation or experimental campaign. The Fourier-coupling construction also provides a starting point for studying other azimuthally periodic profiles through additional magnetic harmonics. Such extensions would require the corresponding sideband couplings and renewed checks of the amplitude ordering, physical validity, and numerical convergence; their stabilizing potential has not been established here. No particular magnetic-circuit implementation or improvement in thruster performance is inferred from the present linear calculation.

\section*{Acknowledgment}

The authors acknowledge support from the National Natural Science Foundation of China (Grant Nos.~52472403 and U25B20235). This work was also partially supported by the Harbin Institute of Technology Kunpeng \& Ascend Center of Cultivation.

\section*{Author Declarations}

\subsection*{Conflict of Interest}

The authors have no conflicts to disclose.

\subsection*{Author Contributions}

\textbf{Yinjian Zhao:} Conceptualization; Methodology; Software; Formal analysis; Validation; Writing -- original draft; Writing -- review and editing; Funding acquisition.

\textbf{Changzheng Hu:} Methodology; Formal analysis; Validation; Writing -- review and editing.

\section*{Data Availability}

The data that support the findings of this study are available from the corresponding author upon reasonable request.

\appendix

\section{\texorpdfstring{Fourier-operator form of \(k_y\) and the meaning of \(\phi(y)\)}{Fourier-operator form of ky and the meaning of phi(y)}}
\label{app:ky_operator}

All appendices use the dimensionless variables of Eq.~\eqref{eq:normalized_variables}. In particular, \(k_y,q,K_B\) are normalized wave numbers, \(y=y^{\rm phys}/\lambda_D\), and \(\ell_B=L_B/\lambda_D\). Thus spatial derivatives, translation distances, Fourier phases, and projection limits use one convention. This appendix explains the formal step
\begin{equation}
  D_0(\omega,\ky;B_0)=0
  \quad\longrightarrow\quad
  D_0(\omega,-\ii\partial_y;B(y))\phi(y)=0.
  \label{eq:appendix_operator_step}
\end{equation}
The step is compact, but it contains three distinct ideas: the Fourier correspondence, the loss of single-\(k_y\) closure in a periodic field, and the interpretation of \(\phi(y)\) as the azimuthal eigenfunction.

\subsection{Fourier correspondence}

Consider a single plane wave,
\begin{equation*}
  \phi(y)=\phi_0 e^{\ii k_y y}.
\end{equation*}
Here \(\phi_0\) is a constant complex amplitude of this trial plane wave.
Taking a derivative gives
\begin{equation*}
  \frac{\partial}{\partial y}e^{\ii k_y y}
  =
  \ii k_y e^{\ii k_y y}.
\end{equation*}
Therefore,
\begin{equation*}
  -\ii\frac{\partial}{\partial y}e^{\ii k_y y}
  =
  k_y e^{\ii k_y y}.
\end{equation*}
Thus the operator \(-\ii\partial_y\) acts on a plane wave exactly as multiplication by its wave number. This is the basic correspondence
\begin{equation*}
  k_y\leftrightarrow -\ii\partial_y.
\end{equation*}
For example,
\begin{equation*}
  k_y\phi\leftrightarrow -\ii\partial_y\phi,
  \qquad
  k_y^2\phi\leftrightarrow(-\ii\partial_y)^2\phi=-\partial_y^2\phi.
\end{equation*}

\subsection{\texorpdfstring{Uniform problem and scalar \(k_y\)}{Uniform problem and scalar ky}}

In a uniform magnetic field, the system is homogeneous in \(y\). A perturbation can be written
\begin{equation*}
  \phi_1(y,t)=\tilde{\phi}e^{\ii(k_y y-\omega t)},
\end{equation*}
where \(\phi_1\) is the first-order electrostatic potential perturbation, \(t\) is time, and \(\tilde{\phi}\) is a constant complex amplitude. Since the coefficients of the linearized problem do not depend on \(y\), a single \(k_y\) is not coupled to any other wave number. Each \(k_y\) can be treated independently, and the stability problem reduces to
\begin{equation*}
  D_0(\omega,k_y;B_0)=0.
\end{equation*}
In this case, \(k_y\) is simply a real parameter chosen by the analyst.

\subsection{Periodic problem and operator form}

For a periodically modulated magnetic field,
\begin{equation*}
  B(y)=B_0[1+\epsB\cos(\kb y)],
\end{equation*}
a single plane wave is no longer closed under the action of the periodic coefficient. The key identity is
\begin{equation*}
  \cos(\kb y)e^{\ii k_y y}
  =
  \frac{1}{2}e^{\ii(k_y+\kb)y}
  +
  \frac{1}{2}e^{\ii(k_y-\kb)y}.
\end{equation*}
Thus the periodic magnetic field couples \(k_y\) to \(k_y+\kb\) and \(k_y-\kb\). Repeated coupling generates \(k_y\pm2\kb\), \(k_y\pm3\kb\), and so on. A single plane wave is therefore not an eigenfunction. The spatial operator must act on the full eigenmode.

\subsection{\texorpdfstring{Meaning of \(\phi(y)\)}{Meaning of phi(y)}}

The function \(\phi(y)\) is the azimuthal eigenfunction of the electrostatic potential perturbation. In the uniform problem one writes
\begin{equation*}
  \phi_1(y,t)=\tilde{\phi}e^{\ii(k_y y-\omega t)}.
\end{equation*}
In the nonuniform periodic problem one writes instead
\begin{equation*}
  \phi_1(y,t)=\phi(y)e^{-\ii\omega t}.
\end{equation*}
If the \(x\) and \(z\) directions are still treated as homogeneous, the perturbation may be written
\begin{equation*}
  \phi_1(x,y,z,t)=\phi(y)e^{\ii(k_xx+k_zz-\omega t)}.
\end{equation*}
In this representation, the \(x\) and \(z\) dependences remain plane waves, while the \(y\) dependence is determined by solving the periodic eigenvalue problem.

\subsection{Connection to the Floquet expansion}

Because \(B(y)\) is periodic, the eigenfunction is expanded as
\begin{equation*}
  \phi(y)=e^{\ii qy}\sum_{\ell=-\infty}^{\infty}\phi_\ell e^{\ii\ell\kb y}
  =
  \sum_{\ell=-\infty}^{\infty}\phi_\ell e^{\ii(q+\ell\kb)y}.
\end{equation*}
The sideband wave numbers are
\begin{equation*}
  k_{y\ell}=q+\ell\kb.
\end{equation*}
On each sideband,
\begin{equation*}
  -\ii\partial_y e^{\ii(q+\ell\kb)y}
  =
  (q+\ell\kb)e^{\ii(q+\ell\kb)y}.
\end{equation*}
Thus, within the Floquet basis, the operator \(-\ii\partial_y\) is equivalent to multiplication by the sideband wave number on each Fourier component.

%\subsection{Caution}
%
%Equation~\eqref{eq:appendix_operator_step} is a semi-local modeling step, not a rigorous full kinetic equation for a nonuniform magnetic field. A full kinetic theory would integrate the linearized Vlasov equation along exact nonuniform electron orbits. The purpose of Eq.~\eqref{eq:appendix_operator_step} is more limited: it keeps the uniform-field kinetic response as a building block and captures the most direct linear effect of periodic modulation, namely Floquet sideband coupling.
%
\section{Periodic coefficients and Floquet waves}
\label{app:floquet_waves}

This appendix explains why eigenfunctions of a linear problem with periodic coefficients can be written in Floquet form. Consider a linear eigenvalue problem in \(y\),
\begin{equation*}
  L(y)\phi(y)=0,
\end{equation*}
where \(L(y)\) denotes the linear operator acting on the azimuthal eigenfunction \(\phi(y)\). The operator contains periodic coefficients. In the present problem this periodicity comes from
\begin{equation*}
  B(y+\ell_B)=B(y),\qquad \ell_B=\frac{2\pi}{\kb}.
\end{equation*}
Therefore,
\begin{equation*}
  L(y+\ell_B)=L(y).
\end{equation*}

Define the translation operator over one period by
\begin{equation*}
  (T_{\ell_B}\phi)(y)=\phi(y+\ell_B).
\end{equation*}
Here \(T_{\ell_B}\) is the operator that translates a function by one magnetic period \(\ell_B\).
Because the coefficients of \(L\) are periodic, applying \(L\) and translating by one period commute:
\begin{equation*}
  LT_{\ell_B}=T_{\ell_B}L.
\end{equation*}
Thus the eigenfunctions of \(L\) can be chosen to be eigenfunctions of \(T_{\ell_B}\) as well:
\begin{equation*}
  T_{\ell_B}\phi(y)=\phi(y+\ell_B)=\mu\phi(y).
\end{equation*}
The scalar \(\mu\) is the eigenvalue of the one-period translation operator. For wave problems it is convenient to write
\begin{equation*}
  \mu=e^{\ii q\ell_B}.
\end{equation*}
Then
\begin{equation*}
  \phi(y+\ell_B)=e^{\ii q\ell_B}\phi(y).
\end{equation*}
This means that the eigenfunction need not be strictly periodic; after one period it may differ from itself by a phase.

Now define
\begin{equation*}
  u_q(y)=e^{-\ii qy}\phi(y).
\end{equation*}
Using the translation relation,
\begin{align*}
  u_q(y+\ell_B)
  &=
  e^{-\ii q(y+\ell_B)}\phi(y+\ell_B)
  \\
  &=
  e^{-\ii qy}e^{-\ii q\ell_B}e^{\ii q\ell_B}\phi(y)
  =
  u_q(y).
\end{align*}
Thus \(u_q(y)\) is periodic with period \(\ell_B\), and
\begin{equation*}
  \phi(y)=e^{\ii qy}u_q(y).
\end{equation*}
Since \(u_q\) is periodic, it can be expanded as
\begin{equation*}
  u_q(y)=\sum_{\ell=-\infty}^{\infty}\phi_\ell e^{\ii\ell\kb y}.
\end{equation*}
Therefore,
\begin{equation*}
  \phi(y)=
  \sum_{\ell=-\infty}^{\infty}\phi_\ell e^{\ii(q+\ell\kb)y}.
\end{equation*}
This proves the Floquet sideband structure used in the main text. The approximation in the present paper lies not in using the Floquet form, which follows from periodic symmetry, but in constructing the periodic operator from the uniform-field kinetic dispersion function.

\section{Projection onto the Floquet basis}
\label{app:projection}

This appendix derives the off-diagonal part of the matrix element. The purpose is to make clear what the indices \(\ell\) and \(m\) mean, where the factor \(1/2\) comes from, and why a sinusoidal magnetic modulation couples only neighboring sidebands at first order.

\subsection{Sideband basis}

Write the Floquet expansion as
\begin{equation*}
  \phi(y)=\sum_{m=-\infty}^{\infty}\phi_m e^{\ii(q+m\kb)y}.
\end{equation*}
Define
\begin{equation*}
  e_m(y)=e^{\ii(q+m\kb)y},\qquad k_{ym}=q+m\kb.
\end{equation*}
Here \(e_m(y)\) is the \(m\)-th Floquet basis function and \(k_{ym}\) is its normalized azimuthal wave number; its dimensional value is \(k_{ym}/\lambda_D\). The index \(m\) labels the input sideband. After the operator acts on \(e_m\), the result is projected onto the output sideband
\begin{equation*}
  e_\ell(y)=e^{\ii(q+\ell\kb)y}.
\end{equation*}
Here \(e_\ell(y)\) is the \(\ell\)-th output basis function. Thus \(D_{\ell m}\) means the contribution from input sideband \(m\) to the projected equation for sideband \(\ell\).

\subsection{Uniform part}

The uniform-field part of the operator acts as
\begin{equation*}
  D_0(\omega,-\ii\partial_y;B_0)e_m
  =
  D_0(\omega,k_{ym};B_0)e_m.
\end{equation*}
The output remains the same sideband. Projection onto \(e_\ell\) contributes only when \(\ell=m\), so the uniform part is diagonal:
\begin{equation*}
  D_{\ell m}^{(0)}
  =
  D_0(\omega,k_{ym};B_0)\delta_{\ell m}.
\end{equation*}

\subsection{First-order magnetic part}

The first-order term is
\begin{equation*}
  \delta B(y)
  \left.
  \frac{\partial D_0(\omega,-\ii\partial_y;B)}{\partial B}
  \right|_{B_0}.
\end{equation*}
For compactness define
\begin{equation*}
  C(-\ii\partial_y)
  =
  \left.
  \frac{\partial D_0(\omega,-\ii\partial_y;B)}{\partial B}
  \right|_{B_0}.
\end{equation*}
Here \(C(-\ii\partial_y)\) is the magnetic-derivative operator obtained from the uniform response.
In the semi-local matrix construction, \(C(-\ii\partial_y)\) acts on the input sideband first. Therefore,
\begin{equation*}
  C(-\ii\partial_y)e_m
  =
  C_m e_m,
  \qquad
  C_m=
  \left.
  \frac{\partial D_0(\omega,k_{ym};B)}{\partial B}
  \right|_{B_0}.
\end{equation*}
The scalar \(C_m\) is this magnetic derivative evaluated on sideband \(m\).

The modulation is
\begin{equation*}
  \delta B(y)=\epsB B_0\cos(\kb y)
  =
  \frac{\epsB B_0}{2}
  (e^{\ii\kb y}+e^{-\ii\kb y}).
\end{equation*}
Multiplying the \(m\)-th sideband gives
\begin{align*}
  \delta B(y)e_m(y)
  &=
  \frac{\epsB B_0}{2}
  (e^{\ii\kb y}+e^{-\ii\kb y})e^{\ii(q+m\kb)y}
  \\
  &=
  \frac{\epsB B_0}{2}
  \left[
  e^{\ii[q+(m+1)\kb]y}
  +
  e^{\ii[q+(m-1)\kb]y}
  \right].
\end{align*}
This is the nearest-neighbor selection rule \(m\to m+1\) and \(m\to m-1\).

\subsection{Projection integral}

The first-order matrix element is
\begin{widetext}
\begin{equation*}
  D_{\ell m}^{(1)}
  =
  \frac{1}{\ell_B}
  \int_0^{\ell_B}
  e^{-\ii(q+\ell\kb)y}
  \left[
  \delta B(y)C(-\ii\partial_y)
  \right]
  e^{\ii(q+m\kb)y}
  \dd y.
\end{equation*}
\end{widetext}
Using the definition of \(C_m\),
\begin{equation*}
  D_{\ell m}^{(1)}
  =
  \frac{C_m}{\ell_B}
  \int_0^{\ell_B}
  e^{-\ii(q+\ell\kb)y}
  \delta B(y)
  e^{\ii(q+m\kb)y}
  \dd y.
\end{equation*}
Substitution of \(\delta B\) gives
\begin{align*}
  D_{\ell m}^{(1)}
  &=
  \frac{\epsB B_0}{2}C_m
  \left[
  \frac{1}{\ell_B}\int_0^{\ell_B}
  e^{\ii(m-\ell+1)\kb y}\dd y
  \right.
  \\
  &\qquad\left.
  +
  \frac{1}{\ell_B}\int_0^{\ell_B}
  e^{\ii(m-\ell-1)\kb y}\dd y
  \right].
\end{align*}
The Fourier orthogonality relation
\begin{equation*}
  \frac{1}{\ell_B}\int_0^{\ell_B}e^{\ii n\kb y}\dd y=\delta_{n0}
\end{equation*}
where \(n\) is any integer and \(\delta_{n0}\) is a Kronecker delta, then gives
\begin{equation*}
  D_{\ell m}^{(1)}
  =
  \frac{\epsB B_0}{2}
  \left.
  \frac{\partial D_0(\omega,k_{ym};B)}{\partial B}
  \right|_{B_0}
  (\delta_{\ell,m+1}+\delta_{\ell,m-1}).
\end{equation*}
This is the off-diagonal part of Eq.~\eqref{eq:matrix_element_b}.

\section{From scalar dispersion relation to matrix dispersion relation}
\label{app:matrix_dispersion}

In a uniform magnetic field, the azimuthal direction is homogeneous and a perturbation can be written as a single plane wave,
\begin{equation*}
  \phi_1(y,t)=\tilde{\phi}e^{\ii(k_y y-\omega t)}.
\end{equation*}
For each prescribed real \(k_y\), the linear stability problem reduces to
\begin{equation*}
  D_0(\omega,k_y;B_0)=0.
\end{equation*}
Each \(k_y\) is independent.

In a periodically modulated magnetic field, one Floquet eigenmode contains a set of sidebands,
\begin{equation*}
  \phi(y)=\sum_{m=-\infty}^{\infty}\phi_m e^{\ii(q+m\kb)y}.
\end{equation*}
The unknown is no longer one scalar amplitude \(\tilde{\phi}\), but the sideband vector
\begin{equation*}
  \phivec=(\ldots,\phi_{-1},\phi_0,\phi_1,\ldots)^{\mathsf{T}}.
\end{equation*}
The projected equation has the form
\begin{equation*}
  \sum_m D_{\ell m}(\omega,q)\phi_m=0,
\end{equation*}
or, after finite truncation,
\begin{equation*}
  \DF(\omega,q)\phivec=0.
\end{equation*}
A nonzero eigenvector requires
\begin{equation*}
  \det\DF(\omega,q;\epsB,\kb)=0.
\end{equation*}
The scalar uniform-field relation is therefore replaced by
\begin{equation*}
  D_0(\omega,k_y;B_0)=0
  \quad\longrightarrow\quad
  \det\DF(\omega,q;\epsB,\kb)=0.
\end{equation*}

\subsection{Minimal three-sideband example}

Keeping only \(m=-1,0,1\), the eigenfunction is
\begin{equation*}
  \phi(y)
  =
  \phi_{-1}e^{\ii(q-\kb)y}
  +
  \phi_0e^{\ii qy}
  +
  \phi_1e^{\ii(q+\kb)y}.
\end{equation*}
The matrix equation has the schematic nearest-neighbor form
\begin{equation*}
  \begin{pmatrix}
    D_{-1}^{(0)} & \mathcal{C}_0 & 0 \\
    \mathcal{C}_{-1} & D_0^{(0)} & \mathcal{C}_1 \\
    0 & \mathcal{C}_0 & D_1^{(0)}
  \end{pmatrix}
  \begin{pmatrix}
    \phi_{-1}\\
    \phi_0\\
    \phi_1
  \end{pmatrix}
  =
  \begin{pmatrix}
    0\\
    0\\
    0
  \end{pmatrix}.
\end{equation*}
Here \(D_m^{(0)}\) denotes the uniform-field response of sideband \(m\), and \(\mathcal{C}_m=(\epsB B_0/2)C_m\) is the nearest-neighbor coupling coefficient, with \(C_m=\left.\partial_B D_0(\omega,k_{ym};B)\right|_{B_0}\) as defined in Appendix~\ref{app:projection}. This example shows directly why the periodically modulated problem is not scalar: even the central sideband is coupled to neighboring sidebands.

\subsection{Practical interpretation}

The practical solution procedure is:
\begin{enumerate}
  \item choose a Floquet quasi-wavenumber \(q\);
  \item choose a finite sideband set \(m=-\orderN,\ldots,\orderN\);
  \item assemble \(\DF(\omega,q)\);
  \item solve \(\det\DF(\omega,q)=0\) for the complex eigenfrequency \(\omega\);
  \item repeat over \(q\) in the first Brillouin zone.
\end{enumerate}
The uniform-field problem asks for roots of a scalar function \(D_0(\omega,k_y)\). The periodic-field problem asks for roots of a determinant because the eigenmode contains multiple coupled sidebands.

\section{Numerical root tracking and quality diagnostics}
\label{app:numerical_diagnostics}

This appendix gives the practical numerical workflow used to support the root-finding procedure described in Sec.~\ref{sec:numerical_method}. It is included to make the distinction between a true Floquet root and a numerical artifact explicit.

For the reference-\(k_z\) period scans, the workflow is as follows. For each modulation pair \((\epsB,K_B)\), sample the first-zone grid and retain a symmetric, padded sideband set. Generate both scalar-frequency branches, previously accepted roots at neighboring \(q\), reflected roots, and amplitude-continuation predictors. Correct them with Eq.~\eqref{eq:floquet_root_vector}, reject non-finite or unconverged candidates, and require \(\sigma_{\min}<10^{-8}\). Remove duplicate roots within \(2\times10^{-7}\). Compare the distinct unstable roots with adaptive contour counts in both rectangles stated in Sec.~\ref{sec:floquet_root_method}; subdivide and add predictors when a count is incomplete. Recheck selected envelope and window roots with the original matrix assembly. These checks validate roots within the stated finite search domains.

Accepted roots are further filtered by their sideband content using the normalized weights in Eq.~\eqref{eq:sideband_weights}. The dominant sideband is the index \(m_\ast\) for which \(w_m\) is largest. The normalized wave number assigned to the root is \(k_{y\ast}=q+m_\ast K_B\); its dimensional value is \(k_{y\ast}^{\rm phys}=k_{y\ast}/\lambda_D\). A root contributes to a specified projected-window diagnostic only when \(k_{y\ast}\) lies inside that window. This assignment distinguishes projected wave-number ranges, rather than providing a unique physical classification of a coupled Floquet mode. It is not applied when constructing the full Bloch-zone envelope. The additional \(k_z\) scan in Sec.~\ref{sec:kz_scan} targets that upper envelope using the finite-strip checks described there.

The final \(q\)-space envelope retains the largest accepted growth at each sampled \(q\). Peak and integrated positive growth use a uniform-field reference computed with the same coverage rules. Report root-count agreement, singular-value residuals, retained \(\orderN\), truncation changes, the sampled \(q\) grid, and \(\chi=\epsB K_B\rho_e\). Numerical root coverage, truncation convergence, finite-grid resolution, and physical model validity are distinct requirements; agreement in one does not establish the others.

\bibliographystyle{aipnum4-2}
\bibliography{references}

@article{Ducrocq2006,
  author  = {Ducrocq, A. and Adam, J. C. and Heron, A. and Laval, G.},
  title   = {High-frequency electron drift instability in the cross-field configuration of Hall thrusters},
  journal = {Physics of Plasmas},
  volume  = {13},
  pages   = {102111},
  year    = {2006},
  doi     = {10.1063/1.2359718}
}

@article{Cavalier2013,
  author  = {Cavalier, J. and Lemoine, N. and Bonhomme, G. and Tsikata, S. and Honore, C. and Gresillon, D.},
  title   = {Hall thruster plasma fluctuations identified as the {$E\times B$} electron drift instability: Modeling and fitting on experimental data},
  journal = {Physics of Plasmas},
  volume  = {20},
  pages   = {082107},
  year    = {2013},
  doi     = {10.1063/1.4817743}
}

@article{Lafleur2016Part1,
  author  = {Lafleur, T. and Baalrud, S. D. and Chabert, P.},
  title   = {Theory for the anomalous electron transport in Hall effect thrusters. I. Insights from particle-in-cell simulations},
  journal = {Physics of Plasmas},
  volume  = {23},
  pages   = {053502},
  year    = {2016},
  doi     = {10.1063/1.4948495}
}

@article{Lafleur2016Part2,
  author  = {Lafleur, T. and Baalrud, S. D. and Chabert, P.},
  title   = {Theory for the anomalous electron transport in Hall effect thrusters. II. Kinetic model},
  journal = {Physics of Plasmas},
  volume  = {23},
  pages   = {053503},
  year    = {2016},
  doi     = {10.1063/1.4948496}
}

@article{Zhou2025PLA,
  author  = {Zhou, Zhijun and Xie, Lihuan and Luo, Xin and Zhao, Yinjian and Yu, Daren},
  title   = {The effect of ion rotational flow on Hall thruster azimuthal instability via two dimensional {PIC} simulations},
  journal = {Physics Letters A},
  volume  = {559},
  pages   = {130899},
  year    = {2025},
  doi     = {10.1016/j.physleta.2025.130899}
}

@article{Lazurenko2008POP,
  author  = {Lazurenko, A. and Coduti, G. and Mazouffre, S. and Bonhomme, G.},
  title   = {Dispersion relation of high-frequency plasma oscillations in Hall thrusters},
  journal = {Physics of Plasmas},
  volume  = {15},
  pages   = {034502},
  year    = {2008},
  doi     = {10.1063/1.2889424}
}

@article{Brown2023PRL,
  author  = {Brown, Zachariah A. and Jorns, Benjamin A.},
  title   = {Growth and Saturation of the Electron Drift Instability in a Crossed Field Plasma},
  journal = {Physical Review Letters},
  volume  = {130},
  pages   = {115101},
  year    = {2023},
  doi     = {10.1103/PhysRevLett.130.115101}
}

@article{Brown2023PRE,
  author  = {Brown, Zachariah A. and Jorns, Benjamin A.},
  title   = {Anomalous cross-field transport in a Hall thruster inferred from direct measurement of instability growth rates},
  journal = {Physical Review E},
  volume  = {108},
  pages   = {065204},
  year    = {2023},
  doi     = {10.1103/PhysRevE.108.065204}
}

@article{Denig2023POP,
  author  = {Denig, A. C. and Hara, K.},
  title   = {Three-dimensional coupling of electron cyclotron drift instability and ion--ion two stream instability},
  journal = {Physics of Plasmas},
  volume  = {30},
  pages   = {032108},
  year    = {2023},
  doi     = {10.1063/5.0122293}
}

@article{Zhao2026PSTReview3DPIC,
  author  = {Zhao, Zhongping and Zhao, Yinjian},
  title   = {A review of 3D particle-in-cell simulations for electron drift instability in Hall thrusters},
  journal = {Plasma Science and Technology},
  year    = {2026},
  volume  = {28},
  pages   = {094002},
  doi     = {10.1088/2058-6272/ae69a9}
}

@article{Chen2026PSTRadialB,
  author  = {Chen, Long and Liu, Yang and Cui, Wanxin and Liu, Miao and Kan, Zichen and Zhang, Quanzhi and Xu, Xuesong and Duan, Ping},
  title   = {2D {PIC} Simulations of Radial Magnetic Field Effects on Electron Drift Instability and Anomalous Axial Transport in Hall Thrusters},
  journal = {Plasma Science and Technology},
  year    = {2026},
  volume  = {28},
  pages   = {094004},
  doi     = {10.1088/2058-6272/ae7257}
}

@article{Zhou2025PSST,
  author  = {Zhou, Zhijun and Luo, Xin and Zhao, Yinjian and Yu, Daren},
  title   = {Impact of azimuthal magnetic field inhomogeneity on Hall thruster high-frequency azimuthal instability via 2D radial--azimuthal {PIC} simulations},
  journal = {Plasma Sources Science and Technology},
  volume  = {34},
  pages   = {105006},
  year    = {2025},
  doi     = {10.1088/1361-6595/ae0af5}
}

@article{Zhou2026PSSTAxial,
  author  = {Zhou, Zhijun and Luo, Xin and Zhao, Yinjian and Yu, Daren},
  title   = {The influence of azimuthally inhomogeneous magnetic field on the high-frequency {Hall} thruster azimuthal instability via {2D} axial-azimuthal {PIC} simulations},
  journal = {Plasma Sources Science and Technology},
  volume  = {35},
  number  = {8},
  pages   = {085002},
  year    = {2026},
  doi     = {10.1088/1361-6595/ae8dfd}
}

@article{Kaganovich2020EXB,
  author = {Kaganovich, I. D. and Smolyakov, A. and Raitses, Y. and Ahedo, E. and Mikellides, I. G. and Jorns, B. and Taccogna, F. and Gueroult, R. and Tsikata, S. and Bourdon, A. and Boeuf, J.-P. and Keidar, M. and Powis, A. T. and Merino, M. and Cappelli, M. and Hara, K. and Carlsson, J. A. and Fisch, N. J. and Chabert, P. and Schweigert, I. and Lafleur, T. and Matyash, K. and Khrabrov, A. V. and Boswell, R. W. and Fruchtman, A.},
  title = {Physics of {$E\times B$} discharges relevant to plasma propulsion and similar technologies},
  journal = {Physics of Plasmas},
  volume = {27},
  pages = {120601},
  year = {2020},
  doi = {10.1063/5.0010135}
}

@article{Janhunen2018MTSI,
  author = {Janhunen, S. and Smolyakov, A. and Sydorenko, D. and Jimenez, M. and Kaganovich, I. and Raitses, Y.},
  title = {Evolution of the electron cyclotron drift instability in two-dimensions},
  journal = {Physics of Plasmas},
  volume = {25},
  pages = {082308},
  year = {2018},
  doi = {10.1063/1.5033896}
}

@article{Boeuf2018PICTheory,
  author = {Boeuf, J. P. and Garrigues, L.},
  title = {{$E\times B$} electron drift instability in {Hall} thrusters: Particle-in-cell simulations vs. theory},
  journal = {Physics of Plasmas},
  volume = {25},
  pages = {061204},
  year = {2018},
  doi = {10.1063/1.5017033}
}

@article{Tsikata2009Scattering,
  author = {Tsikata, S. and Lemoine, N. and Pisarev, V. and Gr{\'e}sillon, D. M.},
  title = {Dispersion relations of electron density fluctuations in a {Hall} thruster plasma, observed by collective light scattering},
  journal = {Physics of Plasmas},
  volume = {16},
  pages = {033506},
  year = {2009},
  doi = {10.1063/1.3093261}
}

@article{Tavassoli2023PICVlasov,
  author = {Tavassoli, A. and {Papahn Zadeh}, M. and Smolyakov, A. and Shoucri, M. and Spiteri, R. J.},
  title = {The electron cyclotron drift instability: A comparison of particle-in-cell and continuum {Vlasov} simulations},
  journal = {Physics of Plasmas},
  volume = {30},
  pages = {033905},
  year = {2023},
  doi = {10.1063/5.0134457}
}

@article{BelloBenitez2025Injection,
  author = {Bello-Ben{\'i}tez, E. and Mar{\'i}n-Cebri{\'a}n, A. and Ahedo, E.},
  title = {Effect of injection conditions on the non-linear behavior of the {ECDI} and related turbulent transport},
  journal = {Physics of Plasmas},
  volume = {32},
  pages = {012307},
  year = {2025},
  doi = {10.1063/5.0238207}
}

@article{Hara2022GradientDrift,
  author = {Hara, K. and Mansour, A. R. and Tsikata, S.},
  title = {Theory of gradient drift instabilities in low-temperature, partially magnetised plasmas},
  journal = {Journal of Plasma Physics},
  volume = {88},
  pages = {905880408},
  year = {2022},
  doi = {10.1017/S002237782200068X}
}

@article{Romadanov2016Nonlocal,
  author = {Romadanov, I. and Smolyakov, A. and Raitses, Y. and Kaganovich, I. and Tian, T. and Ryzhkov, S.},
  title = {Structure of nonlocal gradient-drift instabilities in {Hall} {$E\times B$} discharges},
  journal = {Physics of Plasmas},
  volume = {23},
  pages = {122111},
  year = {2016},
  doi = {10.1063/1.4971816}
}

@article{RamosRipoli2026Kinetic,
  author = {Ramos, J. J. and Ripoli, M.},
  title = {Kinetic theory of drift-gradient modes in a partially magnetised {$E\times B$} plasma},
  journal = {Journal of Plasma Physics},
  volume = {92},
  pages = {E12},
  year = {2026},
  doi = {10.1017/S0022377825101086}
}

@article{Frias2012GradientFluid,
  author = {Frias, W. and Smolyakov, A. I. and Kaganovich, I. D. and Raitses, Y.},
  title = {Long wavelength gradient drift instability in {Hall} plasma devices. {I}. {Fluid} theory},
  journal = {Physics of Plasmas},
  volume = {19},
  pages = {072112},
  year = {2012},
  doi = {10.1063/1.4736997}
}

\end{document}